\documentclass[12pt]{article}
\usepackage[T1]{fontenc}
\usepackage{titlesec, blindtext, color}
\definecolor{gray75}{gray}{0.75}
\newcommand{\hsp}{\hspace{20pt}}
\titleformat{\chapter}[hang]{\Huge\bfseries}{\thechapter\hsp\textcolor{gray75}{|}\hsp}{0pt}{\Huge\bfseries}

\usepackage{lineno}

\usepackage{setspace}
\usepackage[margin=1in]{geometry}
\usepackage{bm}
\usepackage[english]{babel}
\usepackage[latin1]{inputenc}
\usepackage{times}
\usepackage{graphics,graphicx,pgf,natbib,amsmath,amsthm,amssymb,url}
\usepackage{tgbonum}
\usepackage{rotating}
\usepackage{tfrupee}
\usepackage{natbib}
\usepackage{tfrupee}

\usepackage{algorithm}       
\usepackage{algpseudocode}   

\usepackage{xcolor}
\usepackage{subcaption}
\usepackage{hyperref}
\usepackage{tcolorbox}
\usepackage{tikz}
\usetikzlibrary{"arrows", "automata", "backgrounds", "calendar", "chains", "matrix", "mindmap", "patterns", "petri", "shadows", "shapes.geometric", "shapes.misc", "spy", "trees"}

\newcommand{\bfb} {\mbox{\boldmath $\beta$}}

\newcommand{\bfg} {\mbox{\boldmath $\gamma$}}

\newcommand{\bfeta} {\mbox{\boldmath $\eta$}}

\newcommand{\bfl} {\mbox{\boldmath $\lambda$}}

\newcommand{\bfe} {\mbox{\boldmath $\epsilon$}}
\newcommand{\bfep} {\mbox{\boldmath $\varepsilon$}}

\newcommand{\bfm} {\mbox{\boldmath $\mu$}}
\newcommand{\bfS} {\mbox{\boldmath $\Sigma$}}

\newcommand{\bft} {\mbox{\boldmath $\theta$}}

\newcommand{\bfp} {\mbox{\boldmath $\psi$}}

\newcommand{\s}{{\mbox{\boldmath $s$}}}
\newcommand{\f}{{\mbox{\boldmath $f$}}}
\newcommand{\bP}{{\mbox{\boldmath $P$}}}
\newcommand{\p}{{\mbox{\boldmath $p$}}}
\newcommand{\I}{{\mbox{\boldmath $I$}}}

\newcommand{\x}{{\mbox{\boldmath $x$}}}

\newcommand{\X}{{\mbox{\boldmath $X$}}}
\newcommand{\y}{{\mbox{\boldmath $y$}}}
\newcommand{\Y}{{\mbox{\boldmath $Y$}}}

\newcommand{\W}{{\mbox{\boldmath $W$}}}
\newcommand{\w}{{\mbox{\boldmath $w$}}}

\DeclareMathOperator{\argmax}{argmax}

\DeclareMathOperator*{\argmaxB}{argmax}   

\theoremstyle{plain}
\numberwithin{equation}{section}

\newtheorem{theorem}{Theorem}[section]

\newtheorem{remark}{Remark}[section]
\newtheorem{example}{Example}[section]
\newtheorem{experiment}{Experiment}[section]

\title{\textbf{Jacobi Prior: An Alternative Bayesian Method for Supervised Learning}}
\author{
\begin{tabular}{ccc}
    Sourish Das\footnote{Associate Professor at Chennai Mathematical Institute} &  and & Shouvik Sardar\footnote{PhD student at Chennai Mathematical Institute}\\
\end{tabular}
}
\date{28 February, 2026}

\begin{document}
\maketitle

\begin{abstract}
\begin{scriptsize}
   The \emph{Jacobi prior} offers an alternative Bayesian framework for predictive modelling, designed to achieve superior computational efficiency without compromising predictive performance. This scalable method is suitable for image classification and other computationally intensive tasks. Compared to widely used methods such as Lasso, Ridge, Elastic Net, uniLasso, the MCMC-based Horseshoe prior, and non-Bayesian machine learning methods including Support Vector Machines (SVM), Random Forests, and Extreme Gradient Boosting (XGBoost), the Jacobi prior achieves competitive or better accuracy with significantly reduced computational cost. The method is well suited to distributed computing environments, as it naturally accommodates partitioned data across multiple servers. We propose a parallelisable Monte Carlo algorithm to quantify the uncertainty in the estimated coefficients. We establish that the Jacobi estimator is asymptotically close to, and asymptotically equivalent to, the posterior mode under the Jacobi prior.  To demonstrate its practical utility, we conduct a comprehensive simulation study comprising seven experiments focused on statistical consistency, prediction accuracy,  scalability, sensitivity analysis and robustness study. We further present three real-data applications: credit risk modelling using U.S. Small Business Administration (SBA) loan default data, multi-class classification of stars, quasars, and galaxies using Sloan Digital Sky Survey data, and spinal degeneration classification using sagittal MRI scans from the RSNA 2024 Lumbar Spine Degenerative Classification Challenge. In the spine classification task, we extract last-layer features from a fine-tuned ResNet-50 model and evaluate multiple classifiers, including Jacobi-Multinomial logit regression, SVM, and Random Forest. The Jacobi prior achieves state-of-the-art results in recall and predictive stability, especially when paired with domain-specific features. In addition, it substantially outperforms competing methods in runtime by large margins, leading to significant reductions in computational cost. This highlights its potential for scalable, high-dimensional learning in medical image analysis.

All code and datasets used in this paper are available at:
\url{https://github.com/sourish-cmi/Jacobi-Prior/}

\end{scriptsize}

\end{abstract}


\section{Introduction}
In this paper, we present a method for predictive modeling called the `Jacobi prior.' While it was first proposed by \cite{Das_Dey_2006} and \cite{Das2008} for Generalised Linear Models, it was limited to conjugate priors. Here, we expand this idea to nonconjugate priors, especially for modeling nonlinear decision boundaries using the Gaussian process classification method, as discussed in \cite{RasmussenWilliams2005}. We also apply it to multinomial logit regression for $K$-class classification.

A key motivating application for our work is the RSNA 2024 Lumbar Spine Degenerative Classification Challenge \citep{kaggle2024rsna}, which aims to automate the detection and grading of spinal abnormalities from MRI scans. Accurate diagnosis of lumbar degeneration, such as disc narrowing, spinal stenosis, and foraminal narrowing, is critical, particularly among ageing populations where these conditions are a major cause of disability. The task involves high-resolution sagittal spine images with expert-annotated labels across five vertebral levels, posing a complex multi-label classification problem. Traditional deep learning approaches, like end-to-end fine-tuning of ResNet-50, can be effective but require substantial computational resources and specialist expertise, limiting their accessibility in low-resource clinical settings.

This setting naturally calls for scalable methods capable of capturing non-linear decision boundaries in high-dimensional spaces. While multinomial logit regression provides a parametric baseline, it often fails to model such complexity. By imposing a Jacobi prior on the latent space, we enable flexible, data-driven regularisation that better accommodates the intricacies of medical image classification, offering a practical and computationally efficient alternative for large-scale clinical applications.

\subsection{Spine Degeneration Classification}\label{sec_lumber_spine_data}

The RSNA 2024 challenge focuses on detecting degenerative lumbar spinal stenosis (LSS), a common age-related condition. Studies report its prevalence between $1.7\%$ and $13.1\%$, rising significantly with age. The Framingham Study highlights its widespread occurrence and clinical impact \cite{kalichman2009spinal}. The dataset includes thousands of sagittal T2-weighted lumbar spine MRI scans, labelled by radiologists for spinal canal stenosis, subarticular zone narrowing, and neural foraminal narrowing. Each scan is annotated at five vertebral levels (L1-L5) with multi-class severity grades, forming a structured and granular classification task. Given its scale and complexity, conventional Bayesian methods such as MCMC are computationally impractical for training and inference in real-time settings.

The Jacobi prior offers a scalable alternative by producing closed-form estimators for GLMs, avoiding iterative optimization. It enables efficient training from high-dimensional features or deep embeddings, delivering over many fold speed-ups relative to traditional methods. Its analytical tractability also improves model transparency and interpretability, which are essential in clinical applications.

\begin{figure}[ht]
    \centering
    \begin{subfigure}[b]{0.4\textwidth}
        \centering
        \includegraphics[width=\textwidth]{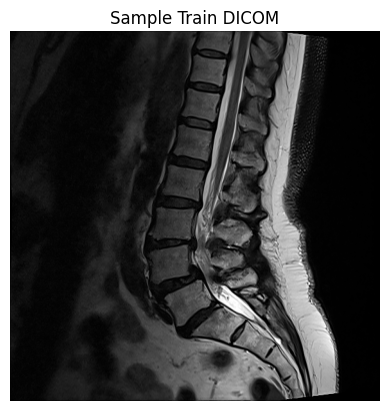}
        \caption{Sagittal view}
        \label{fig:sagittal_view}
    \end{subfigure}
    \hfill
    \begin{subfigure}[b]{0.4\textwidth}
        \centering
        \includegraphics[width=\textwidth]{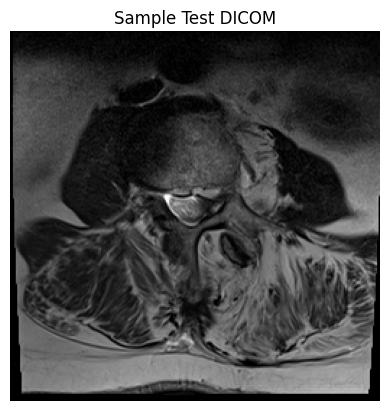}
        \caption{Axial view}
        \label{fig:axial_view}
    \end{subfigure}
    \caption{\scriptsize{Representative MRI slices from the RSNA 2024 Lumbar Spine Degenerative Classification dataset.
    \textbf{(a)} A sagittal T2-weighted DICOM image showing the lumbar vertebrae, intervertebral discs, and spinal canal in longitudinal view. This view is commonly used to assess disc height loss and spinal stenosis.
    \textbf{(b)} An axial DICOM image providing a transverse cross-section of the spinal canal and surrounding structures. This view is useful for identifying lateral recess and foraminal narrowing. Together, these views support comprehensive evaluation of lumbar spine degeneration.}}
    
    \label{fig:rsna_sample_mri}
\end{figure}

\subsubsection*{Dataset Description}

The dataset comprises sagittal and axial T2-weighted lumbar spine MRI scans. Each study includes one or more series of 3D DICOM slices. The goal is to build AI models that classify the severity of three degenerative conditions:
\begin{itemize}
    \item Spinal canal stenosis
    \item Neural foraminal narrowing (labelled separately for left and right sides)
    \item Subarticular zone narrowing (also labelled separately for left and right sides)
\end{itemize}

Annotations are provided at five lumbar levels (L1-L2 to L5-S1) with three ordinal severity grades: normal or mild, moderate, and severe. For the analysis, we binarised these into two categories: ``unaffected" (normal) and ``affected" (moderate or severe).

Figure (\ref{fig:rsna_sample_mri}) displays sagittal and axial MRI slices from two sample images in the dataset. The sagittal view depicts the lumbar vertebrae, intervertebral discs, and spinal canal in longitudinal section, allowing assessment of disc height loss and spinal stenosis. The axial view provides a transverse cross-section of the spinal canal and adjacent structures. Together, these perspectives enable a comprehensive evaluation of lumbar spine degeneration.

\subsection{Motivation for Scalable Bayesian Methods}

Bayesian predictive modelling typically involves placing a prior on regression coefficients. In the Generalised Linear Model (GLM) framework:
\begin{eqnarray} \label{eqn_glm_defn}
	y_i &\sim& f(y_i|\theta_i),~~i=1,2,\cdots,n \nonumber\\
	\theta_i &=& g(\eta_i),~~~\eta_i \in \mathbb{R}\\
	\eta_i &=& \x_i^T\bfb, \nonumber
\end{eqnarray}
where $\x_i=(x_{i1},x_{i2},\cdots,x_{ip})$ is a $p$-dimensional feature vector and the likelihood of the  GLM would be as follows:
$$
L(\bfb|\y,\x) = \prod_{i=1}^{n}f\big(y_i|g(\x_i^T\bfb)\big).
$$
The regular Bayesian method would consider a prior distribution on $\bfb$, say $\pi(\bfb)$.   The 
resulting posterior distribution is
\begin{equation}\label{eqn_gen_post_beta}
    \pi(\bfb|\y,\x) =\frac{L(\bfb|\y,\x)\pi(\bfb)}{\int_{\bfb}L(\bfb|\y,\x)\pi(\bfb)d\bfb}.
\end{equation}
Knowing the posterior distribution is not enough.  We want to evaluate either \textbf{posterior 
	mean}, i.e.,
\begin{equation}\label{eqn_post_mean}
	\mathbb{E}(\bfb|\y,\x)=\int_{\bfb}\bfb \cdot 
 \pi(\bfb|\y,\x) d\bfb,
\end{equation}
or \textbf{posterior mode} of $\bfb$, i.e.,
\begin{equation}\label{eqn_post_mode}
	\hat{\bfb}_{mode} = \argmax_{\bfb}  \pi(\bfb|\y,\x).
\end{equation}
However, both involve high-dimensional integration or optimisation.

In large-scale medical image classification, such as the RSNA 2024 lumbar spine task, the data are high-dimensional and voluminous. Standard Bayesian or machine learning methods that rely on MCMC or optimisation are computationally demanding and often infeasible for real-time use. The Jacobi prior circumvents these issues by providing a closed-form analytical estimator, eliminating the need for MCMC or iterative optimisation. We benchmark it against popular regularisation and Bayesian shrinkage methods, Lasso, Ridge, Elastic Net, and the Horseshoe prior, and find that the Jacobi prior achieves superior predictive accuracy and over 100-fold reductions in computation time.

We present the `Jacobi prior' for Gaussian processes (GP) for the nonlinear classification problem. Standard statistical models, such as logistic regression or discriminant analysis, typically fail to address this problem unless many engineered features are provided. The book by \cite{Goodfellow-et-al-2016}, page 4, introduces the concept of ``deep learning" while criticising standard statistical machine learning models. Here, we demonstrate that the traditional Gaussian process model can effectively handle nonlinear classification using the Jacobi estimator without any difficulties.

Standard methods like Ridge and Lasso solve penalised optimisation problems, interpretable as posterior modes under Gaussian and Laplace priors, respectively \citep{tibshirani1996regression, hastie2009elements}. Their Bayesian versions and the Horseshoe prior rely on MCMC to compute posterior means \citep{carvalho2010horseshoe, Park_Bayes_Lasso_2008}, which is computationally expensive. In contrast, the Jacobi prior provides an exact estimator in closed form.

The Jacobi prior offers several advantages in predictive modelling. First, it provides exact analytical solutions for regression coefficients, removing the need for iterative optimisation or MCMC sampling. This makes it applicable to a broad class of models, including those involving non-linear decision boundaries in classification problems. Second, it is highly scalable, making it particularly effective in cloud-based environments where computational efficiency translates to cost savings, as discussed in \cite{Sandhu_2022}. Third, by reducing computation time and energy consumption, the Jacobi prior contributes to an organisation's sustainability efforts and helps align with ESG compliance standards \cite{Thomas_Benjamin_2023}. Finally, our empirical results demonstrate that models using the Jacobi prior consistently outperform traditional approaches such as Lasso, Ridge, Elastic Net, and Horseshoe priors, both in terms of predictive accuracy and computational speed.

The rest of the paper is structured as follows. Section \ref{sec_Jacobi_prior} introduces the Jacobi prior. Section \ref{sec_sol_Jacobi_prior} derives solutions for logistic, Poisson, and multinomial logit models, including distributed settings. Section \ref{sec_Jacobi_for_GP} extends the approach to Gaussian Processes. Section \ref{sec_simulation_study} presents simulation results, and Section \ref{sec_empirical_study} reports two empirical case studies: classification of Sloan Digital Sky Survey objects and US small business loan approval. Section \ref{sec_spine_classification} presents the detail analysis of results of classification of lumber spine degeneration using ResNet-50 feature representations. Section \ref{sec_conclusion} concludes the paper.

\section{Jacobi Prior}\label{sec_Jacobi_prior}

Often, not knowing the regression coefficients $\bfb$ upfront complicates setting their prior distribution $\pi(\bfb)$ as seen in Equation (\ref{eqn_gen_post_beta}). Nonetheless, we usually possess prior information about natural parameters such as means, proportions, or rate parameters related to the target or dependent variable. In the Jacobi prior method, we define a prior distribution for the natural parameter, for example $\theta_i$, in Equation (\ref{eqn_glm_defn}). This induces a prior distribution on $\eta_i$, as $\theta_i = g(\eta_i)$ is a one-to-one function, using the Jacobian of the transformation. According to the Bayesian paradigm, since $\x_i^T$ is part of the data, it is considered constant, and the prior on $\eta_i$ induces the prior on $\bfb$. In this section, we will analytically elaborate on this idea.


We consider the data set $\mathcal{D}=\big\{(y_i,\x_i)|i=1,2,\cdots,n\big\}$ where the response variable $y_i$ is a function of the $\x_i=(x_{i1},x_{i2},\cdots,x_{ip})$  a $p$-dimension predictor vector for $i^{th}$ sample. The conditional probability model of $y_i$ follows the probability function,
\begin{eqnarray*}
    y_i &\sim& f(y_i|\theta_i),
\end{eqnarray*}
where $\theta_i \in \Theta$ is natural parameters, such that
$$
\theta_i = g(\eta_i), ~~~\eta_i \in \mathbb{R}
$$
is the link function. The $\eta_i$ is modeled as
$$
\eta_i =\x_i^{T}\bfb.
$$
Now $\pi(\theta_i)$ is a prior distribution for $\theta_i$. Then the posterior distribution $\pi(\theta_i|y_i)$ can be modeled as
\begin{eqnarray}
\label{eqn_post_dist_theta}    \pi(\theta_i|y_i)=\frac{f(y_i|\theta_i)\pi(\theta_i)}{\int_{\Theta}f(y_i|\theta_i)\pi(\theta_i)d\theta_i}.
\end{eqnarray}
Since $g()$ is a one-to-one function and we assume that the Jacobian of the transformation exists, the posterior distribution of $\eta_i$ can be obtained using the Jacobian of the transformation. That is
\begin{eqnarray*}
q(\eta_i|y_i)=\pi(g(\eta_i)|y_i)J(\theta_i/\eta_i),
\end{eqnarray*}
where $J(\theta_i/\eta_i)=\Big|\frac{\partial \theta_i}{\partial \eta_i}\Big|=g'(\eta_i)$ is the Jacobian of the transformation. The posterior mode of $\eta_i$ is
\begin{eqnarray}\label{eqn_post_mode_eta}
    \hat{\eta}_i=\argmaxB_{\eta_i} q(\eta_i|y_i) = h(y_i).
\end{eqnarray}
It is important to note that even when we only have knowledge of the posterior distribution $\pi(\theta_i|y_i)$  up to its kernel, we can still derive the posterior mode of $\eta_i$. That is
\begin{eqnarray*}
    \pi(\theta_i|y_i)\propto f(y_i|\theta_i)\pi(\theta_i),
\end{eqnarray*}
can be expressed as
\begin{eqnarray*}
    \pi(\theta_i|y_i)= C f(y_i|\theta_i)\pi(\theta_i),
\end{eqnarray*}
where $C$ is the normalising constant. The resulting posterior distribution of $\eta_i$ is
\begin{eqnarray}\label{eqn_post_dist_eta2}
q(\eta_i|y_i)=Cf(y_i|g(\eta_i))\pi(g(\eta_i))J(\theta_i/\eta_i),
\end{eqnarray}
where the posterior mode of $\eta_i$ can be obtained by optimising the Equation (\ref{eqn_post_dist_eta2}). Under the Kullback-Leibler type loss function, the posterior mode serves as the Bayes estimator for unknown parameters, see \cite{Das2008} and \cite{Das_Dey_2010}.

\subsection{Jacobi prior on $\bfb$}
Now, we need to address a crucial consideration regarding this method. Since we assign a prior to the canonical parameter vector $\bft$, and consequently, we assign a prior to $\bfeta$, does this prior on  $\bfeta$ (which is an $n$-dimension) induce a prior on $\bfb$? Note that $\bfb$ is a $p(<n)$ dimension. The following results presented in \cite{Das2008} demonstrate that it does indeed induce a prior distribution on  $\bfb$.

\begin{theorem}\label{them_prior_beta}
    Suppose $\pi(\bfeta)$ is the prior distribution on $\bfeta$, where  
    $\bfeta=(\eta_1,\eta_2,\cdots,\eta_n).$  Consider $\bfb=A\bfeta$, where $\bfb$ is a 
    $p$-dimension vector and $A$ represents $p$-dimensional linear transformation of 
    $\bfeta$. The prior distribution  $\pi(\bfeta)$ induces a prior distribution on $\bfb$.
\end{theorem}
\begin{proof}\ref{them_prior_beta}\\
We consider the vector $\bfg=(\bfb,\bfp)$, where $\bfb=A\bfeta$ , with $A$ 
represents a $p$-dimensional linear transformation of $\bfeta$. Additionally, $\bfp=C\bfeta$ 
constituting another $(n-p)$-dimensional linear transformation. We specify the prior distribution 
over $\bfg$ as follows
$$
\pi(\bfg)=\pi(\bfb,\bfp)=\pi(A\bfeta,C\bfeta)J(\bfeta/\bfg),
$$
where $\pi(\bfg)=\pi(\bfb,\bfp)$ represents the joint prior distribution over $\bfb$ and $\bfp$. Next, we perform integration over $\bfp$, i.e.,
$$
\pi(\bfb)=\int \pi(\bfb,\bfp) d\bfp.
$$
In other words,  $\pi(\bfb)$ represents the induced prior distribution on $\bfb$, derived from the prior distribution of $\bfeta$. 
\end{proof}
\begin{theorem}\label{them_post_beta}
     Suppose $\pi(\bfeta|\y)$ is the posterior distribution on $\bfeta$.  Considering $\bfb=A\bfeta$, 
     where $A$ 
     represents a $p$-dimensional linear transformation of $\bfeta$.
     The posterior distribution of  $\pi(\bfeta|\y)$ induces a posterior distribution on $\bfb$.
\end{theorem}

\begin{remark}
  The above two theorems from \cite{Das2008} ensure that the prior on the canonical parameter $\bft$ indeed induces a prior on $\bfb$, and the resulting posterior follows from there. 
\end{remark}

\subsection{Jacobi Estimator using the Projection}\label{sec_sub_jacobi_est_projection}
We present the posterior mode in Equation (\ref{eqn_post_mode_eta}), in vector notation as, 
$$
\hat{\bfeta} = (\hat{\eta}_1,\hat{\eta}_2,\cdots,\hat{\eta}_n)=(h(y_1),h(y_2),\cdots,h(y_n))= h(\hat{\y}).
$$ 
The $n$-dimensional vector $h(\y)$ can be uniquely decomposed as
$$
h(\y)=\bP h(\y) + (\I-\bP)h(\y),
$$
where $\bP=\X(\X^{T}\X)^{-1}\X'$ is the full rank model. Note that $\bP$ is the linear transformation matrix representing the orthogonal projection from $n$-dimensional space $\mathbb{R}^n$ onto estimation space $\mathcal{E}(\X)$, while $(\I-\bP)$ represents the orthogonal projection of $\mathbb{R}^n$ onto error space $\mathcal{E}^{\perp}(\X)$. Therefore, 
\begin{eqnarray}\label{eqn_Jacobi_estimate_beta}
\bP h(\y)=\X(\X^{T}\X)^{-1}\X' h(\y)=\X\hat{\bfb},    
\end{eqnarray}
where $\hat{\bfb}=(\X^{T}\X)^{-1}\X' h(\y)$ is the unique least-square estimate of $\bfb$.

\begin{remark}
    In the context of GLM as defined in Equation (\ref{eqn_glm_defn}), there is no error within the classification framework. Thus, the application of a least squares-type estimator as given in Equation (\ref{eqn_Jacobi_estimate_beta}) may appear inappropriate. Despite the GLM formulation positing error-free predictions, the occurrence of misclassification errors prompts a reconsideration of this assumption. The presence of these errors suggests that the model does not explicitly account for all sources of error within the framework of Equation (\ref{eqn_glm_defn}).

    Historically, statisticians have acknowledged such discrepancies, which led to the introduction of the `Anscombe Residual' as discussed in \cite{McCullagh_Nelder_1989}. The concept of the `Anscombe Residual' acknowledges the existence of an error space within the latent variable framework, implying that misclassification errors are essentially manifestations of these latent errors within the predictive model. In this analysis, we recognise that misclassification errors are mapped onto this latent space, and these are captured through the posterior distribution of $\eta$. This perspective allows for a more comprehensive understanding and modeling of error within GLMs.
\end{remark}

\subsection{Parallaisable Monte Carlo Algorithm to Estimate Uncertainty}

We present the parallaisable Monte Carlo algorithm to estimate uncertainty in Algorithm \ref{algo_MC}.
The algorithm provides a Monte Carlo approach for sampling the regression coefficient vector $\bfb$ by leveraging a transformation of latent variables $\theta_i$ drawn independently from their posterior conditionals $\pi(\theta_i \mid y_i)$. At each iteration $r$, the algorithm samples $\theta_i^r$ for all $i = 1, \dots, n$, transforms them via a link function $g$ to obtain $\eta_i^r = g(\theta_i^r)$, and then computes $\beta^r = (X'X)^{-1}X'\eta^r$. The procedure is embarrassingly parallel since each iteration $r$ is independent of others, making it highly scalable and ideal for distributed or GPU-based implementations. Computationally, the algorithm separates the stochastic component (sampling $\theta_i^r$) from the deterministic linear algebra (computing $\beta^r$), and the matrix $X'X$ can be pre-computed and reused, which significantly reduces repeated computation. The quality of the approximation depends on the number of Monte Carlo iterations $N$.

\begin{algorithm}[h]
\caption{Parallelisable Monte Carlo Sampling of $\boldsymbol{\beta}$}\label{algo_MC}
\begin{algorithmic}[1]
\Require Observed data $(X, y)$, simulation size $N$
\For{$r = 1$ To $N$}
  \Statex {\(\triangleright\)} This loop can be parallelised.
  \For{$i = 1$ To $n$}
    \State Sample $\theta_i^{(r)} \sim \pi(\theta_i \mid y_i)$
    \Statex {\(\triangleright\)} As per Eq.~(\ref{eqn_post_dist_theta})
    \State $\eta_i^{(r)} \leftarrow g(\theta_i^{(r)})$
  \EndFor
  \State $\boldsymbol{\eta}^{(r)} \leftarrow \big(\eta_1^{(r)}, \eta_2^{(r)}, \ldots, \eta_n^{(r)}\big)^\top$
  \State $\boldsymbol{\beta}^{(r)} \leftarrow (X^\top X)^{-1} X^\top \boldsymbol{\eta}^{(r)}$
\EndFor
\Ensure Samples $\{\boldsymbol{\beta}^{(r)}\}_{r=1}^N$
\end{algorithmic}
\end{algorithm}

\subsection{Asymptotic Properties of Jacobi Estimator}

The Jacobi estimator proposed in Section~(\ref{sec_sub_jacobi_est_projection}) is based on projection methods in the latent space. We show that the Jacobi estimator is asymptotically close to, and equivalent to, the posterior mode under the Jacobi prior.

\subsubsection*{Model Setup and Notation}
Let $(Y_i, X_i) \in \mathcal{Y} \times \mathbb{R}^p$ be independent data, where $Y_i$ has a distribution in the natural exponential family:
\[
f(y; \theta) = \exp\left\{ y \theta - b(\theta) + c(y) \right\}, \quad \theta \in \Theta \subset \mathbb{R}.
\]
We consider a generalized linear model (GLM) with canonical link:
\[
\theta_i = g(\eta_i),~~~~ \eta_i= \x_i' \bfb.
\]
We adopt a conjugate prior on each $\theta_i$ to obtain a tractable posterior for $\eta_i = \x_i' \bfb$, and define our Jacobi-estimator:
\[
\hat{\bfb}_n = (\X' \X)^{-1} X' \hat{\bfeta}, \quad \text{with } \hat{\eta}_i = \arg\max_{\eta_i} \log \pi(\eta_i \mid y_i),\quad i =1,2,\cdots,n.
\]
The posterior mode under Jacobi prior is:
\begin{eqnarray*}
\tilde{\bfb}_n &=& \arg\max_{\bfb} \big\{\log \pi(\bfb|\y,\X) \big\}\\ 
    & = & \arg\max_{\bfb} \big\{\log \big[L(\bfb|\y,\X)\pi_J(\bfb)\big] \big\},
\end{eqnarray*}
where $\pi_J(\bfb)$ is the induced Jacobi prior as discussed in Theorem (\ref{them_prior_beta}).

\subsubsection*{Assumptions}
\begin{enumerate}
\item[(A1)] The log-likelihood function
$\ell(\y,\bfeta)=\log L(\bfeta|\y,\X)$, where $\bfeta = \X\bfb$ and $\X \in \mathbb{R}^{n \times p}$ has full
column rank. The function $\ell(\y,\bfeta)$ is measurable and twice continuously differentiable
in $\bfeta$.
\item[(A2)] There exists $\epsilon > 0$ such that the prior densities $\pi_J(\bfb)$ and
$\pi(\bfeta)$ satisfy
\[
\pi_J \in C^0(B_\epsilon(\bfb_0)), \quad \pi_J(\bfb) > 0 \ \text{for all } \bfb \in B_\epsilon(\bfb_0),
\]
and
\[
\pi \in C^0(B_\epsilon(\bfeta_0)), \quad \pi(\bfeta) > 0 \ \text{for all } \bfeta \in B_\epsilon(\bfeta_0),
\]
where $B_\epsilon(\bfb_0) := \{ \bfb \in \mathbb{R}^p : \|\bfb - \bfb_0\| < \epsilon \}$ and
$B_\epsilon(\bfeta_0) := \{ \bfeta \in \mathbb{R}^n : \|\bfeta - \bfeta_0\| < \epsilon \}$,
and $C^0(A)$ denotes the space of all continuous functions on the set $A$.

\item[(A3)] The normalized log-likelihood satisfies a uniform law of large numbers:
\[
\sup_{\bfb \in \mathcal{B}}
\left|
\frac{1}{n} \sum_{i=1}^n \ell(Y_i, X_i \bfb)
-
\mathbb{E}\{\ell(\Y, \X \bfb)\}
\right|
\xrightarrow{P} 0 .
\]
\item[(A4)] The function
\[
Q(\bfb) := \mathbb{E}\{\ell(\Y, \X \bfb)\}
\]
has a unique maximizer $\bfb_0$, and the corresponding
$\bfeta_0 := \X \bfb_0$ uniquely maximizes
\[
R(\bfeta) := \mathbb{E}\{\ell(\Y, \bfeta)\}.
\]
\end{enumerate}

\begin{theorem}
Let $\{\hat{\bfb}_n\}_{n \ge 1}$ and $\{\tilde{\bfb}_n\}_{n \ge 1}$ be sequences of random vectors in $\mathbb{R}^p$, then
for every $\varepsilon > 0$ and every $\delta > 0$, there exists an integer
$N(\varepsilon,\delta)$ such that for all $n \ge N(\varepsilon,\delta)$,
\[
\mathbb{P}\!\left( \| \hat{\bfb}_n - \tilde{\bfb}_n \| > \varepsilon \right) < \delta.
\]
\end{theorem}

\begin{proof}
Define the objective function corresponding to the posterior mode in $\beta$-space by
\[
Q_n(\bfb)
=
\frac{1}{n} \sum_{i=1}^n \ell(Y_i, X_i \bfb)
+
\frac{1}{n} \log \pi_J(\bfb),
\]
so that
\[
\tilde{\bfb}_n = \arg\max_{\bfb} Q_n(\bfb).
\]

Similarly, define the objective function for the posterior mode in $\eta$-space by
\[
R_n(\bfeta)
=
\frac{1}{n} \sum_{i=1}^n \ell(Y_i, \eta_i)
+
\frac{1}{n} \log \pi(\bfeta),
\]
and let
\[
\hat{\bfeta}_n = \arg\max_{\bfeta} R_n(\bfeta),
\qquad
\hat{\bfb}_n = (\X' \X)^{-1} \X' \hat{\bfeta}_n .
\]

By Assumption (A2), the prior terms are $O(n^{-1})$ uniformly over compact subsets of $B_\epsilon(\bfb_0)$.

Consequently, $\log \pi_J$ is continuous
on $B_\varepsilon(\bfb_0)$. For any fixed compact set $K \subset B_\varepsilon(\bfb_0)$,
continuity implies that $\log \pi_J$ is bounded on $K$, that is, there exists a finite
constant $M_K < \infty$ such that
\[
\sup_{\bfb \in K} \big| \log \pi_J(\bfb) \big| \le M_K .
\]
It follows that, for every such compact $K$,
\[
\sup_{\bfb \in K} \left| \frac{1}{n} \log \pi_J(\bfb) \right|
\le \frac{M_K}{n},
\]
and therefore
\begin{eqnarray}\label{eqn_vanishing_prior_beta}
\sup_{\bfb \in K} \left| \frac{1}{n} \log \pi_J(\bfb) \right| \longrightarrow 0
\qquad \text{as } n \to \infty.
\end{eqnarray}
Similarly, we can have,  for any fixed compact set $L \subset B_\varepsilon(\bfeta_0)$,
\begin{eqnarray}\label{eqn_vanishing_prior_eta}
\sup_{\bfeta \in L} \left| \frac{1}{n} \log \pi(\bfeta) \right| \longrightarrow 0
\qquad \text{as } n \to \infty.
\end{eqnarray}
Consequently, by Assumption (A3),
\[
\sup_{\bfb \in \mathcal{B}} |Q_n(\bfb) - Q(\bfb)| \xrightarrow{P} 0,
\qquad
\sup_{\bfeta \in \mathcal{H}} |R_n(\bfeta) - R(\bfeta)| \xrightarrow{P} 0,
\]
where $\mathcal{H} = \{ \X\bfb : \bfb \in \mathcal{B} \}$.

By Assumption (A4) and the argmax continuous mapping theorem, it follows that
\[
\tilde{\bfb}_n \xrightarrow{P} \bfb_0,
\qquad
\hat{\bfeta}_n \xrightarrow{P} \bfeta_0 .
\]

Since the mapping $\eta \mapsto (X' X)^{-1} X' \eta$ is linear and continuous,
the continuous mapping theorem implies
\[
\hat{\bfb}_n = (X' X)^{-1} X' \hat{\bfeta}_n \xrightarrow{P} (X' X)^{-1} X'\bfeta_0 =\bfb_0 .
\]

Thus both $\tilde{\beta}_n$ and $\hat{\beta}_n$ converge in probability to the same
limit $\beta_0$, which yields for every $\varepsilon > 0$ and every $\delta > 0$, there exists an integer
$N(\varepsilon,\delta)$ such that for all $n \ge N(\varepsilon,\delta)$,
\[
\mathbb{P}\!\left( \| \hat{\bfb}_n - \tilde{\bfb}_n \| > \varepsilon \right) < \delta.
\]
\end{proof}

\section{Solution for Predictive Models with Jacobi Prior}\label{sec_sol_Jacobi_prior}

\subsection{Logit Regression\label{sec_LR}}

In logistic regression, we have $\y=\{y_1,y_2,\cdots,y_n\}$ are $n$ independent observations from $Binomial(1,p_i)$ with pmf 
$$
f(y_i|p_i)=p_i^{y_i}(1-p_i)^{1-y_i},~~\text{ for all }i=1,2,\cdots,n.
$$
We have the logit link as, $\ln\Big(\frac{p_i}{1-p_i}\Big)=\eta_i,$ and $\eta_i=\x_i^{T}\bfb$ is known as systematic equations. The predictor vector $\x_i'=(x_{i1},x_{i2},\cdots,x_{ip})$ is $p$-dimensional vector, where $\bfb$ is $p$-dimensional vector of unknown coefficients. We consider the conjugate prior for $p_i$, i.e.,
$p_i \sim Beta(a,b),$
then $\ln(\frac{p_i}{1-p_i})=\eta_i\sim \text{Skew-Logistic}(a,b)$ with density as
$$
\pi(\eta_i)=\frac{1}{B(a,b)}\frac{e^{a\eta_i}}{(1+e^{\eta_i})^{a+b}}.
$$
If $a=b=1$, then $\eta_i$ follows standard logistic distribution.  The posterior distribution of $p_i$ 
follows $Beta(y_i+a,1-y_i+b)$, and using the Jacobian of transformation  the posterior distribution of $\eta_i$ is
\begin{eqnarray*}
\pi(\eta_i|y_i)=\frac{1}{B(y_i+a,1-y_i+b)}\frac{e^{(y_i+a)\eta_i}}{(1+e^{\eta_i})^{a+1+b}},    
\end{eqnarray*}
i.e., $\eta_i\sim \text{Skew-Logistic}(y_i+a,1-y_i+b)$. We can determine the posterior mode of $\eta_i$ by taking the derivative of the log-posterior density with respect to $\eta_i$ and then solving the resulting equation by setting it equal to zero. The posterior mode of $\eta_i$ is
\begin{eqnarray}\label{eqn_logit_mode}
\hat{\eta}_i=\ln\Big(\frac{y_i+a}{b+1-y_i}\Big).
\end{eqnarray}
Now using the estimator described in Section (\ref{sec_Jacobi_prior}) in Equation (\ref{eqn_Jacobi_estimate_beta}), we have
$$
\hat{\bfb}=(\X^{T}\X)^{-1}\X^{T} \hat{\bfeta}.
$$
The detailed derivation can be found in \cite{Das_Dey_2006} and \cite{Das2008}.


\noindent \textbf{Vanishing versus non-vanishing prior.}
Assume the joint prior on $\boldsymbol{\eta}=(\eta_1,\ldots,\eta_n)$ factorises as
\[
\pi_n(\boldsymbol{\eta})=\prod_{i=1}^n \pi_{a_n,b_n}(\eta_i),
\qquad
\pi_{a,b}(\eta)=\frac{1}{B(a,b)}\frac{e^{a\eta}}{(1+e^{\eta})^{a+b}}.
\]
Then
\[
\frac{1}{n}\log \pi_n(\boldsymbol{\eta})
= -\log B(a_n,b_n)
+ \frac{1}{n}\sum_{i=1}^n\!\left\{ a_n \eta_i - (a_n+b_n)\log(1+e^{\eta_i}) \right\}.
\]
The term $-\log B(a_n,b_n)$ does not depend on $\boldsymbol{\eta}$ and therefore does not
affect the location of the maximiser of the normalised objective function.

\smallskip
\noindent
\emph{Fixed-shape prior ($a_n=b_n=1/2$).}
In this case the marginal log-density $\log \pi_{1/2,1/2}(\eta)$ is a bounded and non-zero
continuous function with 
\[
\sup_{\eta\in\mathbb{R}}\log \pi_{1/2,1/2}(\eta)=\log(1/(2\pi))<0.
\]
Hence each term $\log \pi_{1/2,1/2}(\eta_i)$ is $O(1)$ on bounded sets, and therefore
\[
\frac{1}{n}\sum_{i=1}^n \log \pi_{1/2,1/2}(\eta_i)
=
O(1).
\]
Consequently, for any localisation sets $L\subset\mathbb{R}^n$ containing vectors with many coordinates
in a fixed bounded interval,
\[
\sup_{\boldsymbol{\bfeta}\in L}
\left|
\frac{1}{n}\sum_{i=1}^n \log \pi_{1/2,1/2}(\eta_i)
\right|
\]
is bounded away from zero. Thus the joint prior contributes an $O(1)$ term to the
normalised objective and does not vanish asymptotically, violating Equation (\ref{eqn_vanishing_prior_eta}). 
The limiting objective function therefore differs from the pure likelihood limit by a non-vanishing deterministic penalty term, which shifts the population maximiser and leads to inconsistency. 
This behaviour is observed in Experiment (\ref{experiment_statistical_consistency}) and visualised in Figure (\ref{fig_simulation_check_consistency_LR}~a). 
The inconsistency persists for any fixed choice $a_n=b_n=\text{constant}$.

\smallskip
\noindent
\emph{Flattening prior ($a_n=b_n=1/n$).}
In this regime,
\[
\log \pi_{1/n,1/n}(\eta)
= -\log B(1/n,1/n) 
+ \frac{1}{n}\eta 
- \frac{2}{n}\log(1+e^{\eta}),
\]
where the $\eta$-dependent part is $O(n^{-1})$ uniformly on bounded sets. Hence
\[
\frac{1}{n}\sum_{i=1}^n \log \pi_{1/n,1/n}(\eta_i)
=
O\!\left(\frac{1}{n}\right)
\longrightarrow 0.
\]
After removing the $\boldsymbol{\eta}$-free constant $-\log B(1/n,1/n)$, the joint prior contribution to the normalised objective is therefore $o(1)$ uniformly on localisation sets where the maximiser concentrates. 
In this sense, the prior becomes asymptotically negligible and does not distort the argmax, satisfying Equation (\ref{eqn_vanishing_prior_eta}) and leading to a consistent estimator. 
This is confirmed in Experiment (\ref{experiment_statistical_consistency}) and visualised in Figure (\ref{fig_simulation_check_consistency_LR}~b).

\smallskip
\noindent
\begin{remark}
It is worth noting that the growth behaviour of the posterior mode 
\[
\hat{\eta}_i = \log\!\left(\frac{y_i+a_n}{1-y_i+b_n}\right)
\]
differs across the two regimes: for $a_n=b_n=1/2$, $\hat{\eta}_i=O(1)$, whereas for $a_n=b_n=1/n$, $\hat{\eta}_i=O(\log n)$. 
However, this divergence of $\hat{\eta}_i$ does not affect consistency of $\hat{\beta}$, since $\hat{\beta}$ depends on averages of $\hat{\eta}_i$ and $\log n / n \to 0$. 
Consistency is determined solely by whether the normalised log-prior term vanishes asymptotically.

\end{remark}


\subsection{Probit Regression\label{sub_sec_probit_reg}}

The Jacobi estimator for logit regression was developed in \cite{Das_Dey_2006} and \cite{Das2008}. Here, we have developed the Jacobi estimator for probit regression, which we will need later in Section (\ref{sec_Jacobi_for_GP}) for modeling the non-linear decision boundary for binary classification with a Gaussian process prior. We have $\y=\{y_1,y_2,\cdots,y_n\}$ are $n$ independent observations from $Binomial(1,p_i)$ with pmf $$f(y_i|p_i)=p_i^{y_i}(1-p_i)^{1-y_i},~~\text{ for all }i=1,2,\cdots,n.$$
We consider the probit link as, $\Phi^{-1}(p_i)=\eta_i$, where $\Phi$ is the cdf of standard Gaussian distribution and $\eta_i=\x_i^{T}\bfb$ is the systematic equations. We consider the conjugate prior for $p_i$, i.e., $p_i \sim Beta(a,b),$ then $\Phi^{-1}(p_i)=\eta_i\sim \text{Skew-Gaussian}(a,b)$ with density as
$$
\pi(\eta_i)=\frac{1}{B(a,b)}\Phi(\eta_i)^{a-1}[1-\Phi(\eta_i)]^{b-1}\frac{1}{\sqrt{2\pi}}e^{-\eta_i^2/2}.
$$
If $a=b=1$, then $\eta_i$ follows standard Gaussian distribution. The posterior distribution of $p_i$ follows $Beta(y_i+a,b-y_i+1)$, and using the Jacobian of transformation  the posterior distribution of $\eta_i$ is
\begin{eqnarray*}
\pi(\eta_i|y_i)=\frac{1}{B(y_i+a,b-y_i+1)}\Phi(\eta_i)^{y_i+a-1}[1-\Phi(\eta_i)]^{(b-y_i+1)-1}\frac{1}{\sqrt{2\pi}}e^{-\eta_i^2/2},   
\end{eqnarray*}
i.e., $\eta_i|y_i\sim \text{Skew-Gaussian}(y_i+a,b-y_i+1)$. The posterior mode for $\eta_i|y_i$ can be obtained as
\begin{equation}
\hat{\eta}_i=\argmax_{\eta\in \mathbb{R}}\pi(\eta_i|y_i).\label{eqn_post_mode_probit_model}    
\end{equation}
Now, employing the estimator outlined in Section (\ref{sec_Jacobi_prior}) as per Equation (\ref{eqn_Jacobi_estimate_beta}), we obtain
\begin{equation}
\hat{\bfb}=(\X^{T}\X)^{-1}\X' \hat{\bfeta}. \label{eqn_jacobi_est_beta_probit_model}   
\end{equation}

\subsection{Poisson Regression}\label{sub_sec_poisson_reg}

Here we briefly present the Jacobi estimator for Poisson regression as presented in  \cite{Das_Dey_2006}. In the next subsection, we will use it to develop the Jacobi estimator for distributed multinomial regression. In Poisson regression, we have $\y=\{y_1,y_2,\cdots,y_n\}$ are $n$ independent observations from $Poisson(\lambda_i)$ with pmf
\begin{eqnarray*}
    f(y_i|\lambda_i)&=&e^{-\lambda_i}\frac{\lambda_i^{y_i}}{y_i!}=\frac{1}{y!}\exp\{y_i\log(\lambda_i)-\lambda_i\},~~~\text{for all }i=1,2,\cdots,n.
\end{eqnarray*}
We have the log link as,
$\ln\big(\lambda_i\big)=\eta_i$ and $\eta_i = \x_i^T\bfb$ is the systematic equation. We consider the conjugate Gamma prior over $\lambda_i$, i.e., $\lambda_i\sim \text{\emph{Gamma}}(a,b),$ with pdf as
\begin{eqnarray*}   \pi(\lambda_i)=\frac{b^{a}}{\Gamma(a)}e^{-b \lambda_i}\lambda_i^{a - 1}.
\end{eqnarray*}
Using the Jacobian of the transformation  $\eta_i=\ln(\lambda_i)$, we have the prior distribution on the $\eta_i$, which follows log-gamma distribution, i.e., $\eta_i\sim \text{\emph{log-Gamma}}(a,b)$
with pdf,
\begin{eqnarray*}
   \pi(\eta_i) &=&\frac{1}{\Gamma(a)}e^{-b \exp\{\eta_i\}}(be^{\eta_i})^{a}.
\end{eqnarray*}
The posterior distribution of $\lambda_i$ follows $\text{\emph{Gamma}}(y_i+a,1+b)$. So the posterior distribution of $\eta_i$ follows $\text{\emph{log-Gamma}}(y_i+a,1+b)$ with pdf,
\begin{eqnarray*}
    \pi(\eta_i|y_i)=\frac{1}{\Gamma(y_i+a)}e^{-(1+b)\exp\{\eta_i\}}\big((1+b)e^{\eta_i}\big)^{(y_i+a)}.
\end{eqnarray*}
We can ascertain the posterior mode of $\eta_i$ by differentiating the log-posterior density with respect to $\eta_i$ and subsequently solving the resulting equation by equating it to zero. The posterior mode of $\eta_i$ is,
\begin{eqnarray}
\hat{\eta}_i=\ln\Big(\frac{y_i+a}{1+b}\Big).\label{eqn_Poisson_jacobi_sol}
\end{eqnarray}
Now, employing the estimator detailed in Section (\ref{sec_Jacobi_prior}) as presented in Equation (\ref{eqn_Jacobi_estimate_beta}), we obtain:
\begin{eqnarray*}
\hat{\bfb} = (\X^{T}\X)^{-1}\X' \hat{\bfeta}.    
\end{eqnarray*}
Next we use it to develop the Jacobi estimator for the distributed multinomial regression.

\subsection{Distributed Multinomial Regression}\label{sec_MLR}

The distributed multinomial regression (DMR) was developed by \cite{Taddy_2015} for text analysis.
Suppose $\Y_i=(Y_{ik})$ is vector of counts in $K$ categories (i.e., $\mathcal{C}_1,\mathcal{C}_1,\cdots,\mathcal{C}_K$), where $k=1,2,\cdots,K$ for $i^{th}$ sample, i.e., $i=1,2,\cdots,n$. The number of counts for $i^{th}$ unit sum up to $m_i=\sum_{k=1}^{K}Y_{ik}$, along with $p$-dimensional predictor vector $\x_i=(x_{i1},x_{i2},\cdots,x_{ip})$. The connection between the category counts $\Y_i$ and the predictor vector $\x_i$ is through multinomial logistic regression (MLR), i.e.,
\begin{eqnarray}
P(\y_i|\x_i,m_i)=Multinomial(\y_i; m_i, \p_i),\label{eqn_mutinomial_model}
\end{eqnarray}
where $\p_i=(p_{i1},p_{i2},\cdots,p_{iK})$, $p_{ik}=\frac{e^{\eta_{ik}}}{\Lambda_i}$, $\Lambda_i =\sum_{k=1}^{K}e^{\eta_{ik}}$, and $\eta_{ik}=\x_i^T\bfb_k$.
\vspace{0.5cm}
Now suppose every $Y_{ik}$ has been drawn independently from the $Poisson(\lambda_{ik})$, where $\lambda_{ik}=e^{\eta_{ik}}$ is the Poisson with intensity $e^{\eta_{ik}}$. The joint likelihood of $\Y_i$ can be expressed as
\begin{eqnarray}\label{eqn_mutinomial_mod2}
P(\Y_i=\y_i)&=&P(\Y_i=\y_i|m_i)\times P(m_i)\nonumber\\
    &=&Multinomial(\y_i; m_i, \p_i)\times Poisson(m_i;\Lambda_i)
\end{eqnarray}
Now $P(\Y_i=\y_i|m_i)=Multinomial(\y_i; m_i, \p_i)$ can be expressed as follows,
\begin{small}
\begin{eqnarray*}
P\Big(Y_{i1}=y_{i1},Y_{i2}=y_{i2},\cdots,Y_{iK}=y_{iK}&|&\sum_{k=1}^{K}Y_{ik}=m_i\Big) \\
&=&\frac{P\Big(Y_{i1}=y_{i1},Y_{i2}=y_{i2},\cdots,Y_{iK}=y_{iK},\sum_{k=1}^{K}Y_{ik}=m_i\Big)}{P(\sum_{k=1}^{K}Y_{ik}=m_i)}\\
    &=&\frac{\prod_{k=1}^{K}e^{-\lambda_{ik}}\frac{\lambda_{ik}^{y_{ik}}}{y_{ik}!}}{e^{-\Lambda_i}\frac{\Lambda_i^{m_i}}{m_i!}}\\
    &=&\frac{m_i!}{y_{i1}!\cdots y_{iK}!}p_{i1}^{y_{i1}}\cdots p_{iK}^{y_{iK}},
\end{eqnarray*}
\end{small}
where $p_{ik}=\frac{\lambda_{ik}}{\Lambda_i}$, $\Lambda_i = \sum_{k=1}^{K}\lambda_{ik}$. In case of the MLR we model the $\p_i$ as follows,
\begin{eqnarray}\label{eqn_softmax}
    p_{ik}=\frac{\exp\{\bfeta_{ik}\}}{\sum_{k=1}^{K}\exp\{\bfeta_{ik}\}},~~~k\leq K,
\end{eqnarray}
where
\begin{eqnarray}
    \bfeta_{ik}=\x_i^T\bfb_k,\label{eqn_DML_Reg}
\end{eqnarray}
where $k=1,2,\cdots,K.$ The full-likelihood of the MLR is computationally expensive, hence \cite{Taddy_2015} presented the DMR while modeling the independent rate parameters 
\begin{eqnarray}\label{Eqn_DMR_Poisson}
    \lambda_{ik}=\exp\{\bfeta_{ik}\}=\exp\{\x_i^T\bfb\}.
\end{eqnarray}
According to the above setup, solving MLR is same as solving $K$ independent Poisson regressions in parallel. Therefore, in each case, we can estimate the $\bfb_k$ using the Jacobi estimator for Poisson regression, as presented in Section (\ref{sub_sec_poisson_reg}).





\subsection{Learning from Partitioned Data  in Distributed Systems}

Often, the sample size ($n$) is quite large and distributed across $M$ systems. We present the partitioned distributed dataset as follows:
\begin{eqnarray*}
    \mathcal{D}=\{\mathcal{D}_1,\mathcal{D}_2,\cdots,\mathcal{D}_M\},
\end{eqnarray*}
where $\mathcal{D}_m=\{(\x_{mi},\y_{mi})|i=1,2,\cdots,n_m\}$, such that $\sum_{m=1}^{M}n_m=n$. Training typical statistical models on such distributed datasets can be challenging. Our Jacobi prior solution can be applied to these distributed datasets. The Jacobi solution for $\bfb$ can be estimated as
\begin{eqnarray*}
    \hat{\bfb}=(\X^T\X)^{-1}\X^T\hat{\bfeta},
\end{eqnarray*}
where $\X$ is $n \times p$ design matrix. Typically, $n$ is very large compared to $p$ and is divided across $M$ systems, denoted as  $\{\X_1,\X_2,\cdots,\X_M\}$. In each system, locally we can calculate the $\X_m^T\X_m$ and $\X_m^T\hat{\bfeta}_m$. Note that $\X_m^T\X_m$ is a $p\times p$ matrix and $\X_m^T\hat{\bfeta}_m$ is a $p\times 1$ vector. These are much smaller and summarised objects. So we can bring these objects into one single computing server and calculate $\X^T\X$ by adding $\X_m^T\X_m$ element-wise, i.e.,
\begin{eqnarray*}
    \X^T\X = \sum_{m=1}^{M}\X_m^T\X_m.
\end{eqnarray*}
Similarly we can also calculate
\begin{eqnarray*}
    \X^T\hat{\bfeta} = \sum_{m=1}^{M}\X_m^T\hat{\bfeta}_m.
\end{eqnarray*}
Then in the master server we can solve $\bfb$ as regular system of equations.

\section{Jacobi Prior for the Gaussian Process Models}\label{sec_Jacobi_for_GP}

\subsection{Binary Classification }

Here, we present Gaussian Process (GP) classification for binary response variables for non-linear decision boundary space or latent space. For the $i^{th}$ sample, $y_i$ is a binary response, with $\x_i$ serving as the corresponding predictor vector. That is
\begin{eqnarray*}
    y_i|\x_i\sim Bernoulli(p_i),~~~i=1,\cdots,n
\end{eqnarray*}
where $p_i = \Phi(\eta_i)$
is the probit link, such that 
\begin{eqnarray}\label{eqn_Binary_GP_regression}
    \eta_i &=&\x_i^T\bfb + w_i +\varepsilon_i,\nonumber\\
    w_i &\sim& \mathcal{GP}(0,\bfS),\\
    \varepsilon_i &\sim& N(0,\sigma^2),\nonumber
\end{eqnarray}
where $w_i$ is the latent random effect and $\varepsilon_i$ is the latent white noise. We consider squared exponential covariance function for the Gaussian process, i.e.,
\begin{eqnarray*}
   \bfS(\x_i, \x_j) = \tau \cdot \exp\left\{ -\rho \, \lVert \x_i - \x_j \rVert \right\},~~i,j=1,2,\cdots,n.
\end{eqnarray*}
The formulation given in Equation (\ref{eqn_Binary_GP_regression}) is recognised as mixed-effect models. In matrix notation, we have
\begin{eqnarray*}
    \bfeta =\X\bfb +\w + \bfe,
\end{eqnarray*}
the marginal distribution of $\bfeta$ is
\begin{eqnarray*}
    \bfeta \sim \mathcal{N}(\X\bfb ~, ~\bfS(\X,\X) + \sigma^2\I_n),
\end{eqnarray*}
where $\bfS(\X,\X) = (\bfS(\x_i, \x_j))$ is the covariance matrix of order $n$. The predictive mean for the test point $\X_0$ would be
\begin{eqnarray*}
    \f_0 = \X_0\bfb +\bfS(\X_0,\X)\Big[\bfS(\X,\X)+\sigma^2\I_n\Big]^{-1}(\bfeta - \X\bfb).
\end{eqnarray*}
The $\bfeta$ and $\bfb$ can be estimated using the Equation (\ref{eqn_post_mode_probit_model}) and Equation(\ref{eqn_jacobi_est_beta_probit_model}) respectively. Hence
the Jacobi estimator for $\f_0$ would be, 
\begin{eqnarray}\label{Eqn_GP_binary_classification_estimator}
    \hat{\f}_0 = \X_0\hat{\bfb} +\bfS(\X_0,\X)\Big[\bfS(\X,\X)+\sigma^2\I_n\Big]^{-1}(\hat{\bfeta} - \X\hat{\bfb}),
\end{eqnarray}
and covariance of $\hat{\f}_0$ is
\begin{eqnarray*}
    Cov(\hat{\f}_0) =\bfS(\X_0,\X)\Big[\bfS(\X,\X)+\sigma^2\I_n\Big]^{-1}\bfS(\X,\X_0).
\end{eqnarray*}

\begin{remark}
    Note that $\sigma$, $\tau$ and $\rho$ are hyper-parameters of the system.
\end{remark}

\begin{remark}
The inverting the matrix $\big[\bfS(\X, \X) + \sigma^2\I_n\big]$ incurs a time complexity of order $\mathcal{O}(n^3)$. The $n$ is the size of the training sample. The solution  requires storing a sizable matrix in the computer's memory, limiting its applicability to moderate-sized datasets. This may pose a significant bottleneck for the solution outlined in Equation (\ref{Eqn_GP_binary_classification_estimator}). This problem can be addressed by sub-sampling technique as proposed by \cite{Das_Roy_Sambasivan_2018}. The algorithm mandates fitting $K$ parallel Gaussian Process solutions to smaller-sized datasets ($n_s$). As a result, the time complexity is $\mathcal{O}(Kn_s^3)$ or $\mathcal{O}(Kn^{3\delta(n)})$. Given that $n$ can be represented as $2^{2^x}$, for $n>2^{2^x}$, the running time is bounded by $\mathcal{O}(Kn^{\frac{3}{x}})$. Consequently, when $n$ is sufficiently large (i.e., $x>3$), the computation becomes sub-linear, as discussed in \cite{Das_Roy_Sambasivan_2018}. when the training dataset size is $n=2^{2^4}=65,536$, the time complexity is $\mathcal{O}(K.n^{3/4})$, where $n^{3/4}=4,096$, resulting in an implementation in sub-linear time. The sub-sampling method works exceptionally well, especially when the sample size ($n$) is very large.    
\end{remark}

\begin{figure}[htbp]
    \centering
    \begin{tabular}{cc}
    \includegraphics[width=0.4\textwidth]{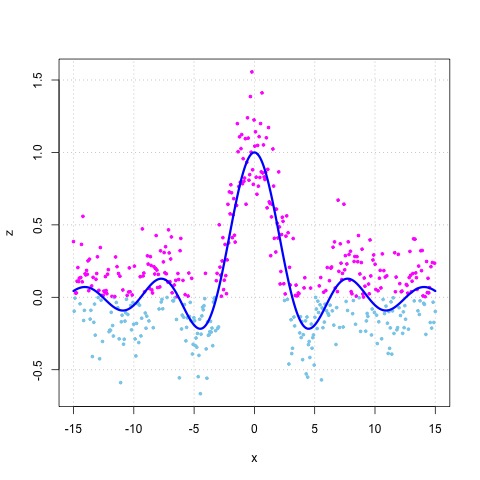}     &  
    \includegraphics[width=0.4\textwidth]{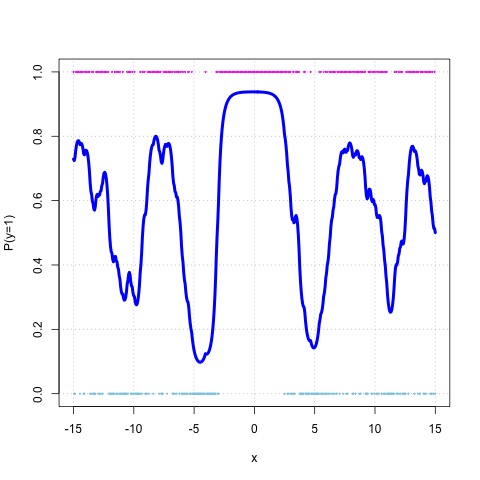}\\
    (a) & (b)\\
     \includegraphics[width=0.4\textwidth]{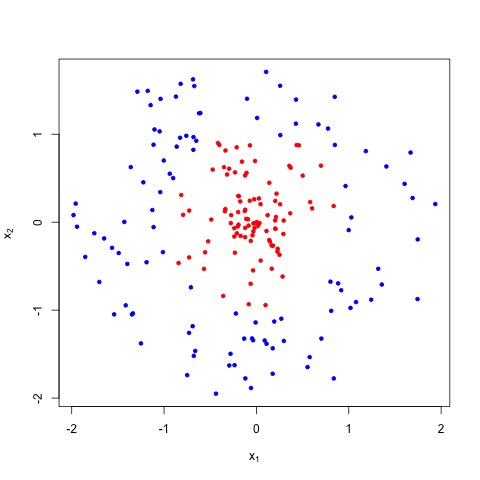}
    &
    \includegraphics[width=0.4\textwidth]{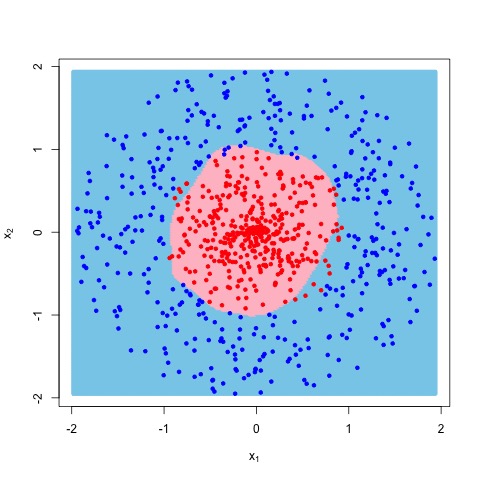}\\
    (c) & (d) \\
    \end{tabular}
    \caption{ \small{Figures (a) and (b) show the simulated data and estimated latent sinc function for Example (\ref{example_sinc_function}). Figures (c) and (d) display the simulated data and estimated circular decision boundary for Example (\ref{example_circular_class}).}}
    \label{fig:GP_classify_EX1}
\end{figure}

\begin{example}\textbf{Latent Sinc Function Learning}\label{example_sinc_function}\\
    Here we consider an example where we simulate the $x$ and $y$ from the sinc function, i.e.,
    \begin{eqnarray*}
        z &=& \frac{\sin{x}}{x} + \varepsilon,\\
        \varepsilon &\sim& N(0,\sigma^2)\\
        y &=& \begin{cases}
            1 & z > 0, \\
            0 & z \leq 0.
        \end{cases}
    \end{eqnarray*}
We consider $x\in (-15,15)$ and simulated $n=500$ samples. Then we pretend that we observed only $x$ and $y$. Using $x$ and $y$, we estimate $\mathbb{P}(y=1) = p$ as a function of $x$ using GP classification with the Jacobi solution for $\bfeta$. In Figure (\ref{fig:GP_classify_EX1}a), we illustrate the simulated values of $x$ and $z$, with points marked in \textcolor{magenta}{magenta} for positive $z$ and \textcolor{cyan}{sky blue} for negative $z$. The \textcolor{blue}{blue} curve represents the sinc function. Moving to Figure (\ref{fig:GP_classify_EX1}b), the same blue curve now represents the estimated probability $\hat{p}=\mathbb{P}(y=1)$ using the Jacobi solution with GP classification. Evidently, the Jacobi solution with GP classification adeptly captures the sinusoidal behavior of the sinc function.
\end{example}


\begin{example}\textbf{Circular Classification}\label{example_circular_class}\\
    In this example, we consider the circular classification problem. The standard statistical models, such as logistic regression or discriminant analysis, generally fail to address the circular classification problem unless we provide many engineered features. The book by \cite{Goodfellow-et-al-2016}, page 4, motivates the idea of ``deep learning," while criticising the standard statistical machine learning models. Here, we demonstrate that the traditional Gaussian process model can address circular classification without any difficulties using the Jacobi estimator. We simulate the data using the following strategy: we generate $x_1$ and $x_2$ using the transformation $x_1=r\sin(\theta)$ and $x_2=r\cos(\theta)$, where $r\sim \text{Uniform}(0,2)$ and $\theta \sim \text{Uniform}(-\pi,\pi)$. We classify the points as \textcolor{red}{red} if $r<1$ and \textcolor{blue}{blue} if $r\geq 1$. We simulate 1000 data points, out of which we keep 200 for training and 800 for testing. We then estimate $\mathbb{P}(y=1) = p$ as a function of $(x_1,x_2)$ using GP classification with the Jacobi solution for $\bfeta$. In Figure (\ref{fig:GP_classify_EX1}c), we showcase the 200 points used for training GP classification with the Jacobi solution. Figure (\ref{fig:GP_classify_EX1}d) depicts the 800 test points and the shaded blue and pink regions illustrate the decision boundary estimated from the 200 training points. The out-of-sample accuracy reaches $96.75\%$.
\end{example}

\subsection{Multi-Class Classification with Distributed Multinomial Regression}

We consider the distributed multinomial regression as presented in Equations (\ref{eqn_mutinomial_mod2},\ref{eqn_DML_Reg},\ref{Eqn_DMR_Poisson}). That is
\begin{eqnarray}\label{eqn_mutinomial_mod3}
P(\Y_i=\y_i)&=&Multinomial(\y_i; m_i, \p_i)\times Poisson(m_i;\Lambda_i),
\end{eqnarray}
where $\p_i=(p_{i1},p_{i2},\cdots,p_{id})$, $p_{ik}=\frac{e^{\eta_{ik}}}{\Lambda_i}$, $\Lambda_i =\sum_{k=1}^{K}e^{\eta_{ik}}$, and $\eta_{ik}=\x_i^T\bfb_k$. As $Y_{ik}$ has been drawn independently from the $Poisson(\lambda_{ik})$, where $\lambda_{ik}=e^{\eta_{ik}}=e^{\x_{ik}^T\bfb_k}$ is the Poisson with intensity. In matrix notation we can express it as
$$
\Y_{k}\sim Poisson(\bfl_{k}), ~~~k=1,2,\cdots,K,
$$
where $\log(\bfl_{k})=\bfeta_{k}.$ There are $k$ equations which can be presented as,
$$
\bfeta =\X\bfb +\W +\bfep,
$$
where $\bfeta$ is $n\times k$ latent response matrix, $\X$ is $n\times p$ feature matrix, $\bfb$ is $p\times k$ regression coefficient matrix, $\W$ is the $n\times k$ random effect matrix and $\bfe$ is $n\times k$ residual matrix. We assume $\W$ is realisation from $K$-variate Gaussian process and $\bfep$ is realisation from $K$-variate standard Normal distribution, i.e.,
\begin{eqnarray*}
\bfep &\sim& N(0,\I_k),\\
\W &\sim& GP(\bf{0},\Sigma).
\end{eqnarray*}
The $\bfeta$ can be estimated as
$$
\hat{\bfeta}=\ln\bigg(\frac{\Y+a}{1+b}\bigg),
$$
the Jacobi estimator for predictive mean for test point $\X_0$ would be
\begin{eqnarray*}\label{Eqn_GP_DML_classification_estimator}
    \hat{\bfeta}_0 = \X_0\hat{\bfb} +\bfS(\X_0,\X)\Big[\bfS(\X,\X)+\sigma^2\I_n\Big]^{-1}(\hat{\bfeta} - \X\hat{\bfb}).
\end{eqnarray*}
Note that once we estimate the $\bfeta$, the prediction would go the same way using the multinomial probability and predicting the class for maximum probability.

\section{Simulation Study}\label{sec_simulation_study}
Here, we present four experiments. The aim of this simulation study is to assess the performance of the Jacobi prior approach in comparison to other widely-used methods, such as MLE, Lasso, Ridge, Elastic Net, and Horseshoe prior, for logistic regression, Poisson regression, and distributed multinomial logistic regression. 

We used a surrogate loss function-based RMSE here. This loss function differs from the classical 0-1 loss function, which measures loss based on strict class mismatch. Instead, the surrogate loss function aims to find the best hyperplane that not only separates the classes but also maximises the distance between the hyperplane and the classes. It measures the deviation between the predicted class membership strength/probability and the two extreme class membership probabilities, 0 and 1. This is a more conservative approach for calculating RMSE. For example, if the predicted strength is 0.70 and the threshold for defining a class is 0.5, the surrogate loss function will give an error value of 0.30, whereas the simple 0-1 loss function will have a 0 loss for this prediction. All the codes for all four experiments are available here: \\\url{https://github.com/sourish-cmi/Jacobi-Prior/}.

\newpage

\begin{experiment}\label{Sim_Ex1}
\textbf{Binary Classification with Logistic Regression}\\
We simulate the data from the true model
\begin{eqnarray}
    \y &\sim& Bernoulli(\p) \nonumber \\
    \p &=& \frac{\exp\{\x^T\bfb\}}{1+\exp\{\x^T\bfb\}} \label{Eqn_logistic_reg}
\end{eqnarray}
We conducted a simulation study with 500 datasets, each comprising a sample size of $n$ with eight predictors. The true parameter vector $\bfb$ was set to $(3,1.5,0,0,2,0,0,0)^T$ and $\sigma$ was set to $3$.  The pairwise correlation between predictors $\x_i$ and $\x_j$ was defined as $\rho(i,j)=0.5^{|i-j|}$ for $i,j=1,2,\cdots,p(=8)$. The design matrix $\X_{n\times p}$ was generated from $\mathcal{N}(\bfm,\bfS)$, where $\bfS=\sigma \rho^{|i-j|}$. Given $\x_i$ for $i=1,2,\cdots,n$ the response variable $y_i$ was generated from Equation (\ref{Eqn_logistic_reg}). We generated a total of $M(=500)$ datasets, denoted as $\mathcal{D}^s=\{\y^s,\X^s_{n\times p}\}$ for $s=1,2,\cdots,M(=500)$ with the true parameter $\bfb=\bfb_0$. For a sample size of $n=100$, the estimates of $\bfb$ from the $s^{th}$ dataset, $\mathcal{D}^s$ were denoted as $\hat{\bfb}^s$. The predicted response $\hat{\p}^s$ was calculated as $\mathbb{E}(\y^s)=\hat{\p}^s=\frac{\exp\{\x^T\hat{\bfb}^s\}}{1+\exp\{\x^T\hat{\bfb}^s\}}$. The root mean square error (RMSE) for the $s^{th}$ data set was computed as
$$
RMSE^s=\sqrt{\frac{1}{n}(\y^s-\hat{\p}^s)^T(\y^s-\hat{\p}^s)}.
$$
We implemented the Jacobi prior using the solution presented in Equation (\ref{eqn_logit_mode}), with $a=b=1/2$ and $m=1$. We implemented the Lasso, Ridge and Elastic Net, using the $5$-fold cross validation method using the \texttt{glmnet} package in \texttt{R}. We implemented the horseshoe prior using \texttt{bayesreg}  package in \texttt{R}. It uses 5000 MCMC samples to estimate the $\bfb$. We implemented Decision Tree, Random Forest, Support Vector Machine, and Extreme Gradient Boosting using the \texttt{rpart}, \texttt{randomForest}, \texttt{e1071}, \texttt{xgboost}, and \texttt{caret} packages in \texttt{R}. The findings are displayed in Table (\ref{Tabl_Sim_Result_Logistic_Reg}). It is worth highlighting that the Jacobi prior demonstrates superior performance compared to other methods, showcasing enhanced predictive accuracy and significantly better runtime complexity. All codes of this experiment are being provided here \url{https://github.com/sourish-cmi/Jacobi-Prior/}.
\end{experiment}

\begin{table}[ht]
\centering
\caption{For the simulation Experiment (\ref{Sim_Ex1}), based on 500 replications, the out sample median RMSE of $\y$, median RMSE of $\beta$ and median time (in microseconds) taken by eleven methods are presented.}
	 \label{Tabl_Sim_Result_Logistic_Reg}
\begin{tabular}{lcc|cc|rr|r}
  \hline
 & RMSE($\y$) & SE & RMSE($\bfb$) & SE & Time & SE & Multiples \\ 
  \hline
  MLE   & 0.69 & 0.01 & 1.97 & 0.28 &  1569 & 0.03 & 10.5 \\ 
  Ridge & 0.61 & 0.00 & 0.97 & 0.00 & 14719 & 0.12 &  98.8 \\ 
  Lasso & 0.66 & 0.00 & 0.58 & 0.01 & 20375 & 0.40 & 136.7 \\ 
  uniLasso & 0.65 & 0.00 & 0.62 & 0.01 & 16343 & 0.10 & 109.7 \\ 
  Elastic Net & 0.66 & 0.00 & 0.63 & 0.01 & 17037 & 0.08 & 114.3 \\ 
  Horseshoe & 0.68 & 0.01 & 0.62 & 0.03 & 2719341 & 5.99 & 18249.2 \\ 
  Decision Tree & 0.64 & 0.00 & NA & NA &    2244 & 0.01 & 15.1 \\ 
  Random forest & 0.58 & 0.00 & NA & NA &   15965 & 0.05 & 107.1 \\ 
  SVM           & 0.63 & 0.00 & NA & NA &    5585 & 0.03 & 37.5 \\ 
  XGB           & 0.65 & 0.01 & NA & NA  & 7766 & 0.02 &  52.1 \\ 
  Jacobi: Probit & 0.55 & 0.00 & 1.26 & 0.00 & 219 & 0.00 & 1.5 \\ 
  Jacobi: Logit & 0.54 & 0.00 & 1.23 & 0.00 & 149 & 0.00 & 1.0 \\ 
   \hline
\end{tabular}

\end{table}

\begin{remark}
    In the first column of the Table (\ref{Tabl_Sim_Result_Logistic_Reg}), we present the out of the sample median RMSE of prediction for different methods calculated over 500 simulated datasets, indicating the predictive performance of Logistic Regression. A lower RMSE indicates better predictive accuracy. The number in second columns are the corresponding standard errors of median RMSE, estimated using the bootstrap with $B=1000$ resampling on the 500 RMSEs. The third column presents median RMSE of the estimates of the $\bfb$. The fifth column shows  the median time taken by each methodology in microseconds across the 500 datasets. \emph{It is noteworthy that the Jacobi prior outperforms other methods, exhibiting both superior predictive accuracy and more efficient time complexity.} In the last column, we  provide the time taken by different methods relative to the Jacobi prior solution. The Jacobi  prior is over 200 times faster than any method using \texttt{glmnet} and approximately 18 times faster than Fisher's scoring-based MLE.
\end{remark}

\begin{remark}
In the current landscape of cloud computing, time is often equated with monetary value, as cloud services are billed according to usage duration. The Jacobi prior stands out for providing accurate predictions and doing so much faster. This increased efficiency not only saves money for cloud users but also helps organisations reduce their environmental impact. Reducing operational time and energy use supports ESG goals, offering both cost savings and environmental benefits.
\end{remark}

\begin{remark}
    It is counter intuitive that $RMSE(\bfb)$ is higher for Jacobi prior compare to other methods. Note that, the Jacobi prior is elicited directly on the mean parameter $\theta_i = \mathbb{E}(y_i)$, rather than on the regression coefficients $\bfb$. Consequently, the method is inherently focused on improving the estimation of $\theta_i$, which directly governs the predictive performance on $y_i$, rather than on recovering $\bfb$ itself. This could explain why we observe strong predictive accuracy despite relatively poorer estimation of $\bfb$.
\end{remark}

\begin{experiment}
    \textbf{Statistical Consistency of Jacobi  Estimator for Logistic Regression}\\Here we consider the setup of Experiment (\ref{Sim_Ex1}). In this experiment, we want to check if the Jacobi estimator of $\bfb$ for the logistic regression model presented in Equations (\ref{Eqn_logistic_reg}) is statistically consistent. We considered the increasing sample size from $n=200$ to $5000$. For each sample size, we ran the experiment exactly the same as Experiment (\ref{Sim_Ex1}). When we consider $a=b=1/2$, for all choices of $n$, we calculated the RMSE of $\bfb$ and plotted the results in Figure (\ref{fig_simulation_check_consistency_LR}a), which shows that the Jacobi estimator's RMSE is constant. 
    It implies the Jacobi estimator is inconsistent when $a=b=1/2$, i.e., when $a$ and $b$ are constant. Next we consider the Jacobi prior when $a=b=1/n$; calculated the corresponding RMSE of $\bfb$ for different choices of $n$ and plotted the results in Figure (\ref{fig_simulation_check_consistency_LR}b). When we choose $a=b=\frac{1}{n}$, the RMSE of Jacobi estimator decreases with increasing $n$, like MLE. It implies that both bias and variance tend to zero with increasing sample size. Hence, the experiment indicates that the Jacobi estimator is statistically consistent when $a=b=\frac{1}{n}$.
    \begin{figure}[htbp]
        \centering
         \centering
        \begin{tabular}{cc}
        \includegraphics[width=0.32\textwidth]{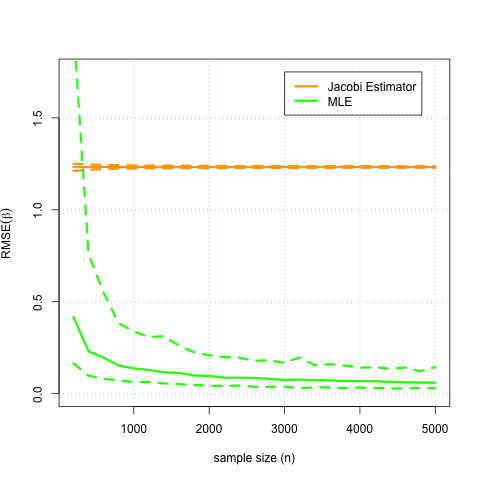}
        &
        \includegraphics[width=0.32\textwidth]{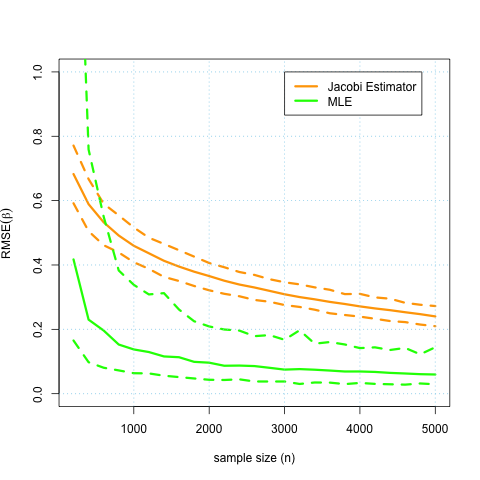}
        \\
        (a)&(b)
        \end{tabular}
        \caption{ Figure presents the RMSE of $\bfb$ of Logistic Regression with increasing sample size ($n$), when (a) $a=b=1/2$; and (b) $a=b=1/n$}
    \label{fig_simulation_check_consistency_LR}
    \end{figure}
\end{experiment}

\begin{experiment}\label{experiment_poisson_ref}
\textbf{Poisson Regression for Count Data}\\
    We simulate the data from the true model
\begin{eqnarray}
    \y &\sim& Poisson(\lambda) \nonumber \\
    \lambda &=& \exp\{\x^T\bfb\}. \nonumber \label{Eqn_Poisson_reg}
\end{eqnarray}
We performed a simulation study with 500 datasets, each having a sample size of $n$ and eight predictors. The true parameter vector $\bfb$ was set to $(0.3,0.15,0,0,0.2,0,0,0)^T$ and $\sigma$ was set to $1$. The pairwise correlation between predictors $\x_i$ and $\x_j$ was setup as in the Example (\ref{Sim_Ex1}). For a sample size of $n=100$, the estimates of $\bfb$ from the $s^{th}$ dataset, $\mathcal{D}^s$ were denoted as $\hat{\bfb}^s$. The predicted response $\hat{\lambda}^s$ was calculated as $\mathbb{E}(\y^s)=\hat{\lambda}^s=\exp\{\x^T\hat{\bfb}^s\}$. We implemented the Jacobi prior using the solution presented in Equation (\ref{eqn_Poisson_jacobi_sol}), with $a=b=1$ and $m=1$. We implemented Lasso, Ridge, and Elastic Net using the 5-fold cross-validation method with the \texttt{glmnet} package in \texttt{R}. We implemented Decision Tree, Random Forest, Support Vector Machine, and Extreme Gradient Boosting using the \texttt{rpart}, \texttt{randomForest}, \texttt{e1071}, \texttt{xgboost}, and \texttt{caret} packages in \texttt{R}. 
The findings are displayed in Table (\ref{Tabl_Sim_Result_Poisson_Reg}). It is worth noting that the Jacobi prior demonstrates superior performance compared to other methods, showcasing enhanced predictive accuracy and more efficient time complexity.
\end{experiment}

\begin{table}[ht]
	\centering
\caption{Median RMSE and Time Complexity of Experiment (\ref{experiment_poisson_ref})}.
\label{Tabl_Sim_Result_Poisson_Reg}
\begin{small}
\begin{tabular}{lcr|rr}  \hline 
\textbf{Methods}	&  RMSE($\y$) & SE & Time & Multiplier \\   \hline
MLE & 1.33 & 0.01 & 887.87 & 11.67\\   
Ridge & 1.22 & 0.02 & 13482.45 & 177.27 \\   
Lasso & 1.23 & 0.01 & 10830.40 & 142.40 \\   
Elastic Net & 1.23 & 0.01 & 11435.39 & 150.36 \\   
Horseshoe  & 1.22 & 0.01 & 3647811.06 & 47962.47 \\
Decision Tree  & 1.34 & 0.01 & 2318.50 & 30.48 \\
Random Forest  & 1.22 & 0.01 & 30022.98 & 394.75 \\
SVM  & 1.26 & 0.01 & 1911.88  & 25.14 \\
XGB  & 1.35 & 0.01 & 12373.57   & 162.69 \\
Jacobi Prior & \textbf{1.16} & 0.01 & 76.06 & 1.00 \\    
\hline
\end{tabular}    
\end{small}

\end{table}

In the first column of Table (\ref{Tabl_Sim_Result_Poisson_Reg}), we present the Median RMSE  of different methods calculated over  500 datasets, indicating the predictive performance of Poisson regression. A lower RMSE  indicates better predictive accuracy. The second column displays the standard deviation of RMSE across these 500 datasets. The third column shows the median time taken by each methodology in microseconds. Notably, the Jacobi prior outperforms other methods, showcasing both superior predictive accuracy and more efficient time complexity. In the fourth column, we present the time taken by different methods relative to the Jacobi prior solution. The Jacobi prior is more than 150 times faster than any \texttt{glmnet} methods and about 15 times faster than Fisher's scoring-based MLE.

\newpage

\begin{experiment}\textbf{K-Class Classification with Distributed Multinomial Logistic Regression}\label{experiment_DMR_simulation}\\
    We simulate the data from the model
\begin{eqnarray}
    \y &\sim& Multinomial(\s,\p) \nonumber \\
    \p &\sim& (p_1,p_2,\cdots,p_K) \nonumber \\
    p_k &=& \frac{\exp\{\x^T\bfb_k\}}{\sum_{k=1}^{K}\exp\{\x^T\bfb_k\}}, \nonumber \\
    \x &=& \{x_1,x_2,\cdots,x_P\}.\nonumber \label{Eqn_Poisson_reg2}
\end{eqnarray}
We consider number of class as $K=4$, number of features as $P=3$, sample size $n=50$ and simulated 100 such data sets; where we simulated $\bfb$ for each data sets using following strategy,
\begin{eqnarray*}
    \bfb_{(P+1)\times K}&\sim& \mathcal{N}(0,1),\\
    \beta_{j,k} &=& \begin{cases}
      \beta_{j,k} & \text{with probability 0.5},   \\
      0 & \text{with probability 0.5}.
    \end{cases}
\end{eqnarray*}
where $j=1,2,\cdots,P$, $k=1,2,\cdots,K$. The strategy induces $50\%$ sparsity in the solution space. Once the coefficient matrix $\bfb$ is simulated it is treated as true parameter, denoted as $\bfb_0$. The predictors $\x$ is simulated from \texttt{uniform(0,1)} and response is being $\y$ from \texttt{Multinomial}($\s,\p$), where $\s \sim Poisson(\lambda=10)$. We applied the Lasso, Ridge, and Elastic Net methods using the 5-fold cross-validation technique, utilising the \texttt{glmnet} package in \texttt{R}. Similarly, we employed the \texttt{bayesreg} package in \texttt{R} to implement the horseshoe prior. This method uses 12000 MCMC samples to estimate the $\bfb$ and we implemented it for all 100 datasets.
\end{experiment}

\begin{table}[ht]
\centering
\caption{Median RMSE and Run-Time Complexity of Experiment (\ref{experiment_DMR_simulation})}
\label{Tabl_simu_result_DMR}
\begin{small}
\begin{tabular}{lccrr}
  \hline
\textbf{Methods} & Median RMSE & Sd of RMSE & Time & Multiple \\ 
  \hline
\emph{Full Multinomial}   &&&&\\
\emph{Regression}               &&&&\\
~~~~ Elastic Net & 0.72 & 0.35 & 62437.53 & 931.96 \\ 
~~~~ Ridge & 0.72 & 0.33 & 61874.15 & 923.56 \\ 
~~~~ Lasso & 0.74 & 0.36 & 70528.51 & 1052.73 \\ 
\emph{Distributed Multinomial}   &&&&\\
\emph{Regression}               &&&&\\
~~~~ Elastic Net & 0.71 & 0.38 & 46859.03 & 699.43 \\ 
~~~~ Ridge & 0.74 & 0.38 & 55206.06 & 824.02 \\ 
~~~~ Lasso & 0.71 & 0.38 & 46835.54 & 699.08 \\ 
~~~~ Horseshoe & 0.74 & 0.33 & 9855186.00 & 147101.94 \\ 
~~~~ Jacobi Prior & \textbf{0.62} & 0.32 & 66.99 & 1.00 \\ 
   \hline
\end{tabular}    
\end{small}
\end{table}

In the first column of the Table (\ref{Tabl_simu_result_DMR}), we report the median RMSE for various methods, calculated across 100 datasets. This serves as an indicator of the predictive efficiency in DMLR regression, where a lower RMSE signifies enhanced predictive accuracy. The second column illustrates the standard deviation of the RMSE over these datasets. The median duration each method, measured in microseconds, is depicted in the third column. Notably, the Jacobi prior demonstrates exceptional performance, excelling in both predictive accuracy and computational efficiency. The fourth column compares the time efficiency of different methods to that of the Jacobi prior. Remarkably, the Jacobi prior is over 600 times quicker than any \texttt{glmnet} method and horseshoe prior.

\begin{experiment}\textbf{Sensitivity Analysis of Jacobi Prior for Logistic Regression}\label{experiment_sensitivity_simulation}\\
Here we consider the setup of Experiment (\ref{Sim_Ex1}). In this experiment, we want to check if the Jacobi estimator of $\bfb$ for the logistic regression model presented in Equations (\ref{Eqn_logistic_reg}) is sensitive to choice of hyperparameters $a$ and $b$. The heatmap presented in Figure (\ref{fig:sensitivity_analysis}) illustrates how the predictive performance of the estimator varies with different values of the tuning parameters $a$ and $b$, which are used to transform the response variable before estimating regression coefficients. For each $(a, b)$ pair on a fine grid, the model is trained on the training data and evaluated on the test data using Root Mean Squared Error (RMSE) between predicted probabilities and actual labels. The colour gradient in the heatmap, ranging from brown (low RMSE) to blue (high RMSE), reveals that smaller values of both $a$ and $b$ generally lead to better model performance. This sensitivity analysis helps identify optimal parameter combinations that minimise prediction error and highlights how the estimator's effectiveness depends on the choice of these transformation parameters. This observation, supports the Remark (\ref{remark:logistic_reg_hyper_param_choices}).
\begin{figure}[h]
    \centering
    \includegraphics[width=0.6\linewidth]{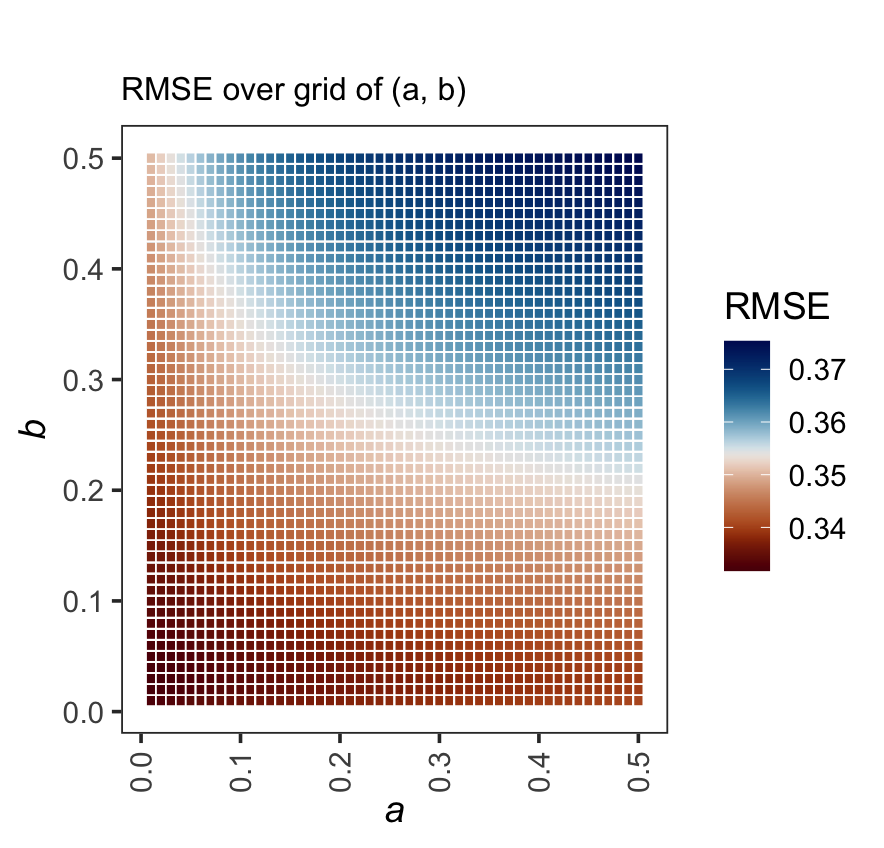}
    \caption{Sensitivity Analysis of Jacobi Estimator for logistic regression. RMSE over grid of (a,b) is being presented, where a and b are hyper-parameters.}
    \label{fig:sensitivity_analysis}
\end{figure}
\end{experiment}

\begin{experiment}\textbf{Robustness Study of Jacobi Prior for Logistic Regression}\label{experiment_robust_logistic}\\
In this experiment, we retain the same setup as in Experiment (\ref{Sim_Ex1}), with one key modification: after generating the data, we randomly select $10\%$ of the instances and flip the corresponding values of $y_i$ to introduce label contamination. The results are presented in Table (\ref{Tabl_Sim_Result_Logistic_Reg_Robust}). As expected, the performance of most methods deteriorates under contamination. Nevertheless, the Jacobi prior continues to outperform competing methods, demonstrating robust predictive accuracy along with substantially higher computational efficiency.
\end{experiment}
\begin{table}[h]
\centering
\caption{Results of Experiment (\ref{experiment_robust_logistic}) assessing robustness under 10\% response label flipping. Based on 500 replications, the table reports the median out-of-sample RMSE for \(\mathbf{y}\), the median RMSE for \(\boldsymbol{\beta}\), and the median computation time (in microseconds) across eleven methods.}
	 \label{Tabl_Sim_Result_Logistic_Reg_Robust}
\begin{small}
    \begin{tabular}{lcc|cc|rr|r}
  \hline
 & RMSE($\y$) & SE & RMSE($\bfb$) & SE & Time & SE & Multiples \\ 
  \hline
MLE & 0.70 & 0.01 & 1.97 & 0.28 & 1572.49 & 0.05 & 10.27 \\ 
  Ridge & 0.62 & 0.00 & 0.97 & 0.00 & 14360.55 & 0.31 &  93.82 \\ 
  Lasso & 0.66 & 0.00 & 0.59 & 0.02 & 19301.41 & 0.56 & 126.10 \\ 
  Elastic Net & 0.66 & 0.00 & \textbf{0.61} & 0.01 & 16330.00 & 0.09 & 106.69\\ 
  Horseshoe & 0.68 & 0.01 & 0.64 & 0.03 & 2638596.30  & 9.13 & 17238.43 \\ 
  Decision Tree & 0.64 & 0.00 & NA & NA & 2220.87 & 0.02 & 14.51 \\ 
  Random forest & 0.59 & 0.00 & NA & NA & 15639.54 & 0.07 & 102.18 \\ 
  SVM & 0.64 & 0.00 & NA & NA & 5588.53 & 0.03 & 36.51 \\ 
  XGB & 0.66 & 0.00 & NA & NA  & 7687.57 & 0.03 &  50.22 \\ 
  Jacobi: Probit & 0.56 & 0.00 & 1.26 & 0.00 & 222.92 & 0.00 & 1.46 \\ 
  Jacobi: Logit & \textbf{0.55} & 0.00 & 1.23 & 0.00 & 153.06 & 0.00 & 1.00 \\ 
   \hline
\end{tabular}
\end{small}
\end{table}

\begin{experiment}\label{experiment_robust_poisson_reg}\textbf{Robustness Study of Jacobi Prior for Poisson Regression}\\
This experiment follows the same setup as Experiment (\ref{experiment_poisson_ref}), with a single modification: after generating the response data, we randomly contaminate 10\% of the responses ($y$) by replacing them with draws from a Poisson distribution with \(\lambda = 20\), thereby introducing significant outliers. The corresponding results are reported in Table (\ref{Tabl_Sim_Result_Poisson_Reg_Robust}). As anticipated, the predictive performance of all methods declines under such heavy contamination. Nonetheless, the Jacobi prior remains one of the best-performing approaches, alongside SVM, exhibiting strong robustness in prediction while maintaining superior computational efficiency.
\end{experiment}

\begin{table}[h]
	\centering
\caption{Median RMSE and computation time for Experiment (\ref{experiment_robust_poisson_reg}) with 10\% contamination introduced by replacing responses with samples from a Poisson distribution with \(\lambda = 20\).}
\label{Tabl_Sim_Result_Poisson_Reg_Robust}
\begin{small}
\begin{tabular}{lcr|rr}  \hline 
\textbf{Methods}	&  RMSE($\y$) & SE & Time & Multiplier \\   \hline
MLE & 6.43 & 0.13  & 981.57  & 13.09\\   
Ridge & 5.82 & 0.09 & 13556.00  & 180.79 \\   
Lasso & 5.80 & 0.11 &  10877.97  &  145.07 \\   
Elastic Net & 5.79 & 0.10 & 11821.03 &  157.65 \\   
Horseshoe  & 5.88 & 0.07 & 3686996.10 & 49171.33 \\
Decision Tree  & 6.37 & 0.08 &  2292.04   &  30.57 \\
Random Forest  & 5.75 & 0.08 & 31502.49  &  420.13 \\
SVM  & \textbf{5.72} & 0.10  &  1886.01  &   25.15 \\
XGB  & 6.04 & 0.11  &  12930.04  &  172.44  \\
Jacobi Prior & \textbf{5.72} & 0.09 &   74.98  &    1.00  \\    
\hline
\end{tabular}    
\end{small}
\end{table}

\newpage

\section{Empirical Study}\label{sec_empirical_study}
\subsection{Galaxies, Quasars, and Stars: Three Class Classification}

In the field of astronomy, the classification of stars, galaxies, and quasars using machine learning is vital for gaining insights into the universe's diverse components (see \cite{Clarke_2020}). This classification, based on spectral characteristics, holds significant importance. The dataset under consideration comprises 100,000 space observations collected through the Sloan Digital Sky Survey (SDSS) \cite{Abdurrouf_2022}. Each observation is defined by seven distinct features, complemented by an additional class label that designates it as a star, galaxy, or quasar. This dataset is tailored to streamline the classification of stars, galaxies, and quasars, making use of their distinctive spectral properties.  The descriptions of the seven features of the SDSS data sets is presented in the Table (\ref{tab_digital_sky_data_features}).
\begin{table}[htbp]
    \centering
     \caption{Description of the features of the SDSS datasets.}
    \label{tab_digital_sky_data_features}
    \begin{tabular}{c|c|l} \hline
       & Feature Name   &  Description\\ \hline
      1 & alpha   & Right Ascension angle (at J2000 epoch) \\
      2& delta & Declination angle (at J2000 epoch) \\
      3& u & Ultraviolet filter in the photometric system\\
      4& g & Green filter in the photometric system \\ 
      5& r & Red filter in the photometric system \\
      6& i & Near Infrared filter in the photometric system \\
     7& z & Infrared filter in the photometric system \\
      \hline
    \end{tabular}
   
\end{table}

\begin{figure}[h]
    \centering
    \begin{tabular}{cc}
       \includegraphics[width=0.4\textwidth]{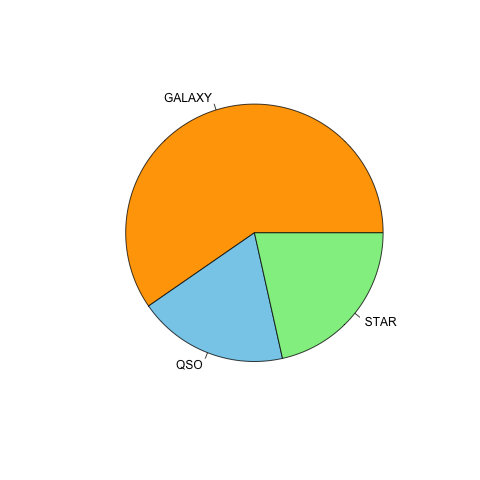}  &  
       \includegraphics[width=0.4\textwidth]{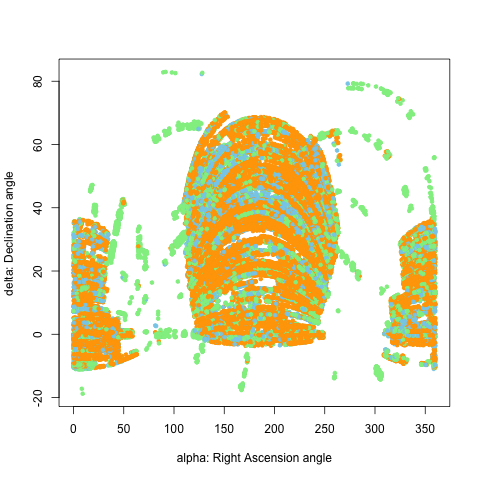} \\
       (a) & (b) \\
       \includegraphics[width=0.4\textwidth]{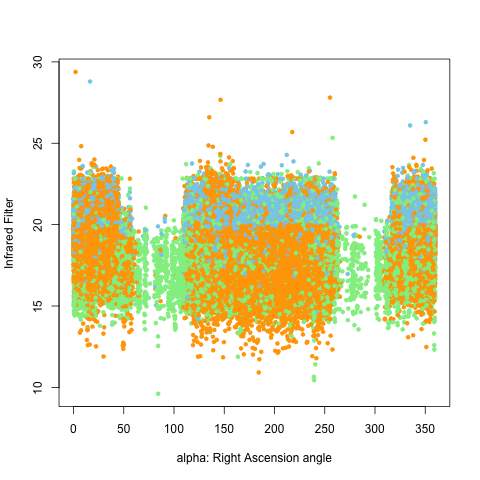} &
       \includegraphics[width=0.4\textwidth]{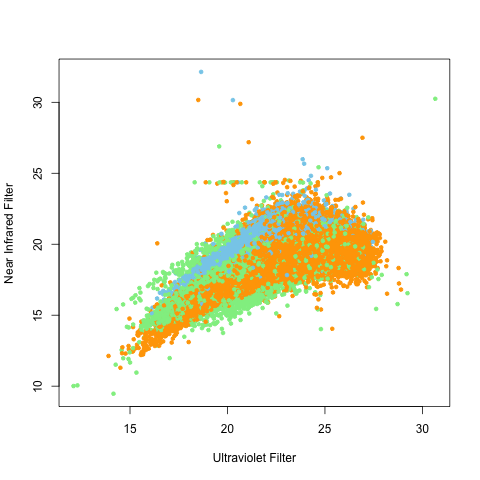}
       \\
       (c) & (d) \\
    \end{tabular}
    \caption{A graphical presentation of SDSS data to visualise different features and the type of astronomical object.}
    \label{fig_sdss_eda}
\end{figure}
The dataset is split into train and test sets. Seventy percent of the data is randomly selected for training, while the remaining ($30\%$) is kept for testing. Exploratory data analysis (EDA) is conducted using the training dataset, and a few graphs are presented in Figure (\ref{fig_sdss_eda}). Detailed EDA can be found at the following GitHub link: \url{https://github.com/sourish-cmi/Jacobi-Prior/}. The pie chart in Figure (\ref{fig_sdss_eda}a) illustrates the distribution of galaxies, quasars, or stars in the training dataset. Figure (\ref{fig_sdss_eda}b) shows the scatter plot between alpha (i.e., right ascension angle) and delta (i.e., declination angle), indicating a highly non-linear relationship between the variables. Figure (\ref{fig_sdss_eda}c) presents the scatter plot between alpha and infrared filter. The graph indicates that when the right ascension angle (i.e., alpha) is between 60-120 or 260-320, the chance that the object is a star is very high. Similarly, Figure (\ref{fig_sdss_eda}d) presents the scatter plot between ultraviolet filter and near-infrared filter, indicating highly non-linear relationships. It is evident that stars are significantly different from galaxies and quasars.

We consider the task of 3-class classification, where the classes are namely stars, galaxies, and quasars. We implement the Distributed Multinomial Regression with Poisson model, incorporating Ridge, Lasso, and Elastic Net regularisation, along with the Jacobi Prior. In addition, we also implement the regular Multinomial Regression model using the \texttt{glmnet} package in \texttt{R}.  In Table (\ref{tabl_accuracy_sdss_star_classification}), we present the out-of-sample accuracy and run-time (in seconds) of different methods. The out-of-sample accuracy of the Jacobi prior and the Lasso penalty are almost the same. However, in terms of run-time, the Jacobi prior is nearly 500 times faster than the Lasso penalty.

\begin{table}[ht]
\centering
\caption{Out-sample accuracy and runtime complexity comparison of different methods for Multinomial regression on the SDSS star classification dataset.}
\label{tabl_accuracy_sdss_star_classification}
\begin{tabular}{lrrr}
  \hline
  \textbf{Methods} & Accuracy &  Time & Multiple \\ 
  \hline
 Distributed Multinomial Regression  &&&\\ 
 ~~~ with Jacobi Prior & \textbf{73.05} & 0.69 & 1.00 \\ 
 ~~~ with Ridge Penalty & 61.67 & 13.06 & 18.89 \\ 
 ~~~ with Lasso Penalty & \underline{72.94} & 348.58 & 504.28 \\ 
 ~~~ with Elastic Net & 72.71 & 376.63 & 544.86 \\ \hline
 Full Multinomial Regression  &&&\\ 
 ~~~ with Ridge Penalty & 62.83 & 65.21 & 94.33 \\ 
 ~~~ with Lasso Penalty & 70.97 & 575.74 & 832.93 \\ 
 ~~~ with Elastic Net  &  72.48 & 641.84 & 928.53 \\ 
   \hline
\end{tabular}
\end{table}

\subsection{Loan Approved or Denied}
In this study, we employed a rich dataset obtained from the U.S. Small Business Administration (SBA) \citep{SBA_Loan_Data_Kaggle, Li_2018}, comprising 899,164 samples. This dataset captures a diverse range of scenarios, including notable success stories of startups benefiting from SBA loan guarantees, as well as instances of small businesses or startups facing defaults on their SBA-guaranteed loans. Our primary objective in analysing this dataset is to predict the probability of default. Upon reviewing the dataset, the central question emerges: should a loan be granted to a specific small business? This decision hinges on a thorough evaluation of various factors, necessitating a nuanced exploration of the loan approval or denial criteria. In this study we considered five features, described in Table (\ref{tab_feature_description_loan_approval_datasets}).

\begin{table}[ht]
    \centering
       \caption{Description of the features of the US SBA Loan Approval Dataset}
    \label{tab_feature_description_loan_approval_datasets}
    \begin{tabular}{c|l|l} \hline
       & Feature Name  &  Description\\ \hline
       1  & New & 1 = Existing business, 2 = New business \\
       2 & RealEstate & If backed by real estate \\
       3 & DisbursementGross  &  Amount disbursed \\
       4 & Term     & Loan term in months \\
       5 & NoEmp    & Number of business employees \\ \hline
    \end{tabular}
\end{table}

\noindent We implemented simple logistic regression using the \texttt{glm} function in \texttt{R}. Additionally, we implemented regularisation methods such as Lasso, Ridge, and Elastic Net using the \texttt{glmenet} package in \texttt{R}. We also implemented the Jacobi prior with four different techniques. To choose the hyperparameters $a$ and $b$, we considered $a=b=0.5$ (Jeffrey's prior for $p$); $a=6/10$, and $b=4/10$, an ad-hoc choice. Furthermore, we implemented a stochastic search for $a$ and $b$, and finally, we employed a Bayesian optimisation method to estimate $a$ and $b$. The prediction in the loan default problem entails two types of errors with different consequences. Along with the accuracy of the prediction, we also estimate the utility of the prediction. We define the utility of the prediction for a single loan in the following Table (\ref{tab_utility_of_loan_default_prediction}). We calculate the utility of all the loans and add them up.
\begin{table}[ht]
    \centering
     \caption{Utility of the prediction. The $V$ is the Gross Disbursement amount.}
    \label{tab_utility_of_loan_default_prediction}
    \begin{tabular}{l|cc} \hline
                & Denie  & Approve  \\ \hline
       Default  &  $V*0.1$   &   $-V*0.7$  \\
       No Default & $-V*0.1$ &   $V*0.5$ \\ \hline
    \end{tabular}
   
\end{table}

Out-sample accuracy, utility of prediction and runtime complexity comparison of different methods are presented in Table (\ref{tabl_accuracy_of_loan_default_pred_dataset}). The outcomes from the dataset on loan approval or denial by the USA Small Business Association reveal that employing Jacobi Prior with Stochastic Optimisation leads to higher accuracy, enhanced utility, and an average speed improvement of more than 200 times compared to conventional solutions such as Lasso, Ridge, and Elastic Net.

\begin{table}[ht]
\centering
\caption{Out-sample accuracy, utility of prediction and runtime complexity comparison of different methods for Logistic regression, on the US SBA loan dataset.}
\label{tabl_accuracy_of_loan_default_pred_dataset}
\begin{tabular}{lrrrr}
  \hline
Method & Accuracy & Utility & Time & Multiples\\ 
  \hline
MLE & 82.74 & 19.31 & 1.49 & 16.56\\ 
  Lasso & 82.70 & 19.32 & 20.06 &  222.88\\ 
  Ridge & 82.72 & 19.46 & 38.73 & 430.33 \\ 
  Elastic Net & 82.68 & 19.32 & 21.85 & 242.78\\ 
  Jacobi Prior: a=b=0.5 & 82.35 & 19.16 & 0.09 & 1.00\\ 
  Jacobi Prior: a=6/10, b=4/10 & \textbf{83.41} & 19.31 & 0.09 & 1.00\\ 
  Jacobi Prior: Stochastic Optim & 83.04 & \textbf{19.82} & 0.11 & 1.22\\ 
  Jacobi Prior: Bayes Optim & 83.19 & \textbf{19.84} & 19.75 & 219.44\\ 
   \hline
\end{tabular}
\end{table}

\section{Classification of Spine Degeneration Using ResNet-50 Feature Representations}
\label{sec_spine_classification}

\subsection{Stratified Sampling Design for Supervised Learning}

To construct balanced and representative datasets for supervised learning, we employed a stratified sampling strategy that accounted for both clinical condition labels and anatomical levels. The dataset comprised 24{,}546 unique sagittal MR images, each annotated with one or more instances of spinal conditions, resulting in a total of 48{,}692 image-condition instances.

Each instance was categorised as either \emph{affected} or \emph{unaffected} based on the reported severity score. Instances with severity zero were labelled as unaffected (serving as background), whereas those with non-zero severity were marked as affected, indicating the presence of a degenerative spinal condition.

Two disjoint subsets were created for model development:
\begin{itemize}
    \item A training set consisting of 4,592 condition-level samples drawn from 4,042 distinct image IDs,
    \item A validation set comprising 482 samples from 473 unique image IDs.
\end{itemize}

Both subsets were constructed to include 20\% unaffected and 80\% affected samples. Within each subset, sampling was stratified across anatomical levels and diagnostic categories to preserve both class diversity and anatomical representativeness. Sampling was performed with a fixed random seed to ensure reproducibility. Validation samples were drawn exclusively from instances not appearing in the training set to avoid data leakage.

\subsection{Anatomical Localisation via Faster R-CNN}

\begin{figure}[h]
    \centering
    \includegraphics[width=0.32\linewidth]{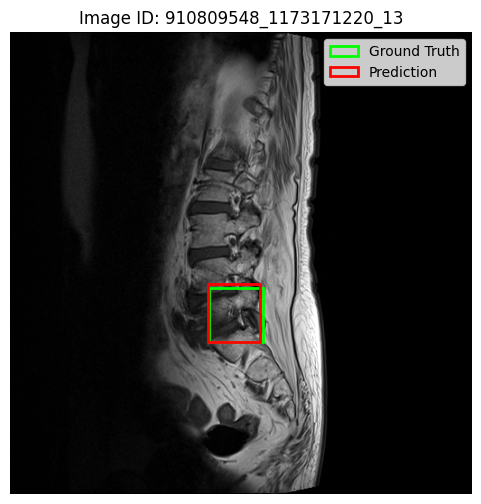}
    \includegraphics[width=0.32\linewidth]{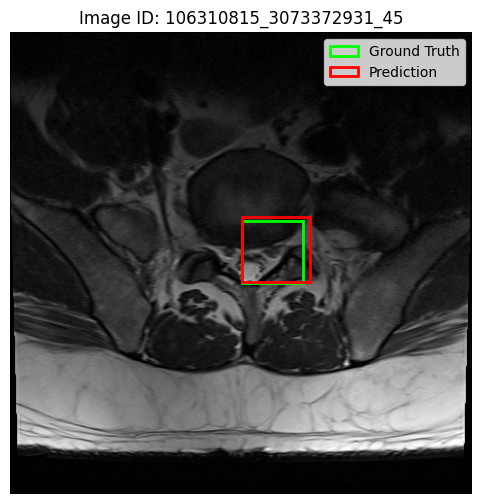}
    \includegraphics[width=0.32\linewidth]{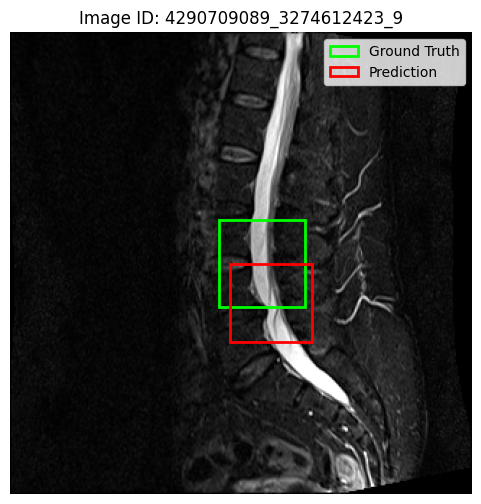}
    \caption{Visual comparison of predicted (red) and ground truth (green) bounding boxes on three MRI scans. The first two examples show good alignment between prediction and ground truth, while the third image demonstrates misaligned bounding box. }
    \label{fig:bounding_box_results}
\end{figure}

Prior to classification, we trained a 25-class Faster R-CNN model to localise spinal conditions at specific vertebral levels. Figure~\ref{fig:bounding_box_results} presents qualitative results on three representative cases, showing predicted bounding boxes (red) and ground truth annotations (green). In the first two examples, the predicted bounding boxes are well-aligned with the ground truth, capturing the target regions with high spatial fidelity. The third image, however, reveals less overlapping predictions, reflecting reduced precision. These visualisations illustrate both the strengths and limitations of the detection module in complex anatomical contexts.

\subsection{Classification Using Deep Feature Representations}

We evaluated multi-label classification performance under three modelling frameworks:

\begin{enumerate}
    \item Direct classification using predictions from ResNet-50 without any data-specific fine-tuning.
    \item Traditional classifiers (Random Forest, Support Vector Machine, and Jacobi-Multinomial Logit) trained on last-layer features extracted from a pretrained ResNet-50.
    \item The same classifiers trained on features extracted from a ResNet-50 model retrained on our lumbar spine dataset, using 5 epochs and 144 batch size.
\end{enumerate}

Performance was evaluated on a per-image basis using four standard metrics:
(i) exact set match between predicted and true condition sets,
(ii) full recall of the true set (i.e., $\text{true set} \subseteq \text{predicted set}$), and
(iii) mean Intersection over Union (IoU) between predicted and ground truth condition sets.

\begin{table}[h!]
\centering
\caption{Classification performance across model configurations. Metrics are computed over 482 test samples.}
\label{tab:evaluation_summary}
\begin{small}
    \begin{tabular}{lcccr}
\hline
\textbf{Method} & \textbf{Exact Match} & \textbf{$\geq 100\%$ Recall} & \textbf{Mean IoU} & \textbf{Time (Seconds)}\\[0.5ex]
\hline
\multicolumn{5}{l}{\textit{Direct ResNet-50 Predictions}} \\
Pretrained ResNet-50 & 0.0053 & 0.0080 & 0.0067 & -- \\[0.5ex]
Retrained ResNet-50  & 0.0775 & \textbf{0.4118} & -- & 127.53 \\[1ex]
\hline
\multicolumn{5}{l}{\textit{Traditional ML on Pretrained Features}} \\
Random Forest         & 0.0374 & 0.1283 & 0.0789 & 150.54\\[0.5ex]
Support Vector Machine & 0.0428 & 0.1765 & 0.1045 & 18.23 \\[0.5ex]
Jacobi-Multinomial Logit & 0.0588 & \textbf{0.3155} & \textbf{0.1700} & 1.15 \\[1ex]
\hline
\multicolumn{5}{l}{\textit{Traditional ML on Retrained Features}} \\
Random Forest         & \textbf{0.0561} & 0.4652 & 0.2472 & 108.46 \\[0.5ex]
Support Vector Machine & 0.0535 & \textbf{0.4759} & 0.2495 & 3.16 \\[0.5ex]
Jacobi-Multinomial Logit & 0.0348 & 0.4439 & 0.2212 & 1.38 \\[0.5ex]
\hline
\end{tabular}
\end{small}
\end{table}

\subsection{Interpretation of Results}

Several insights emerge from the comparison:

\begin{itemize}
    \item \textbf{Fine-tuning the feature extractor is essential.} The pretrained ResNet-50 performs poorly when used directly or when its features are fed into conventional classifiers. Retraining the network on domain-specific data improves full recall from 6.4\% to over 40\% and boosts IoU substantially, while exact match remains limited due to the inherent challenge of multi-label recovery across all models.
    
    \item \textbf{Jacobi-Multinomial Logit exhibits robustness with limited supervision.} Among the models trained on pretrained features, Jacobi-Logit achieves the highest full recall and mean IoU, outperforming both Random Forest and SVM. This result highlights the Jacobi model's ability to extract signal from weakly informative features via structured regularisation.
    
    \item \textbf{Retrained features enhance all models.} With features extracted from the retrained ResNet-50, all classifiers demonstrate significant improvements in recall and IoU. Random Forest achieves the highest IoU and exact match, although SVM and Jacobi-Logit remain competitive.
    
    \item \textbf{Exact set match remains low.} Across all configurations, exact match is generally under 5\%, reflecting the difficulty of predicting the exact condition set correctly for each image without producing too many false positives. Nevertheless, high recall and IoU indicate clinically meaningful partial accuracy.
\end{itemize}

These results highlight the value of combining deep feature learning with structured, probabilistic models for complex classification tasks in medical imaging. In particular, the Jacobi-Logit framework offers a promising balance between computational scalability, uncertainty quantification, and predictive performance in high-dimensional, complex labeled settings.

\section{Concluding Remarks}\label{sec_conclusion}

This study introduces the Jacobi prior as an alternative Bayesian method designed to address the computational challenges of traditional techniques. Our experiments demonstrate that Jacobi method achieves superior predictive accuracy compared with established methods such as Lasso, uniLasso, Ridge, Elastic Net, Horseshoe prior, Decision Tree, SVM, Random Forest, and XGBoost, while running over 100 times faster.

The Jacobi prior is versatile, with successful applications in Generalised Linear Models, Gaussian process regression, and classification. It is well suited for distributed computing, efficiently handling partitioned data located on servers across continents. Notably, the Jacobi estimator maintains strong predictive performance, suggesting that modelling the mean parameter $\theta_i$ directly, can enhance prediction,  a phenomenon that opens up a scope for further theoretical investigation.

It's substantial runtime advantage also carries environmental significance: reduced computational demand translates into a smaller carbon footprint, aligning with organisations' long-term ESG commitments. Thus, the Jacobi prior offers both technical and sustainability benefits.

Our simulation and empirical analyses confirm its effectiveness across varied contexts, including medical image classification, credit risk estimation, and astronomical classification. The accompanying open-source code and datasets in our GitHub repository promote reproducibility and further exploration.

In summary, the Jacobi prior stands out as a robust, efficient, and widely applicable tool for Bayesian inference, offering a compelling alternative to traditional methods and paving the way for advances in predictive modelling across diverse domains.

\bibliographystyle{chicago}
\bibliography{biblio_list}




\end{document}